\documentclass[letterpaper,11pt]{article}
\pdfoutput=1
\usepackage{jcappub}
\usepackage{etoolbox}
 \makeatletter
    \patchcmd{\maketitle}{\@fpheader}{}{}{}
    \makeatother
\usepackage{threeparttable} 
\usepackage{color}
\usepackage{epstopdf}
\usepackage{natbib}
\usepackage{mathrsfs}
\usepackage{subcaption}
\usepackage{tikz-cd}
\usepackage{mdwlist}
\usepackage{dsfont}
\usepackage{natbib}
\usepackage{amssymb}
\usepackage{amsthm}
\usepackage{graphicx}
\usepackage{amsmath}
\usepackage{url}
\usepackage{bm}
\usepackage{epstopdf}
\usepackage{color}
\usepackage{enumitem}
\usepackage{dsfont}
\usepackage{bbm}
\usepackage{mathrsfs}
\usepackage[utf8]{inputenc}

\newcommand{\di}{\partial}

\newcommand{\be}{\begin{equation}}
\newcommand{\ee}{\end{equation}}
\newcommand{\de}{\mbox{d}}

\newcommand{\pa}{\partial}

\newcommand{\RicThree}{{}^{(3)}\mathcal{R}}

\newcommand{\cmmnt}[1]{}

\numberwithin{equation}{section}

\begin{document}

\title{The singularity in mimetic Kantowski-Sachs cosmology}

\author{Marco de Cesare, Sanjeev S.\ Seahra, and Edward Wilson-Ewing}

\affiliation{Department of Mathematics and Statistics, University of New Brunswick \\ Fredericton, NB, Canada E3B 5A3}

\emailAdd{marco.de\_cesare@unb.ca}
\emailAdd{sseahra@unb.ca}
\emailAdd{edward.wilson-ewing@unb.ca}

\abstract{The dynamics of the vacuum Kantowski-Sachs space-time are studied in the so-called limiting curvature mimetic gravity theory.  It is shown that in this theory the vacuum Kantowski-Sachs space-time is always singular.  While the departures from general relativity due to the limiting curvature mimetic theory do provide an upper bound on the magnitude of the expansion scalar, both its rate of oscillations and the magnitude of the directional Hubble rates increase without bound and cause curvature invariants to diverge.  Also, since the radial scale factor does not vanish in finite (past) time, in this particular theory the Kantowski-Sachs space-time cannot be matched to a null black hole event horizon and, therefore, does not correspond to the interior of a static and spherically symmetric black hole.}

\date{\today}

\maketitle

\section{Introduction}

Limiting curvature mimetic gravity was originally introduced in order to provide a solution to the initial singularity in cosmology \cite{Chamseddine:2016uef}. The theory is a modification of the original mimetic theory \cite{Chamseddine:2013kea} by means of a (multi-valued) function depending on the d'Alembertian of the mimetic scalar field which introduces an upper bound to the expansion, with this upper bound a free parameter in the theory. In cosmological space-times, the expansion is simply the mean Hubble rate and for the spatially flat Friedmann-Lema{\^i}tre-Robertson-Walker (FLRW) and Bianchi~I space-times, this modified gravity theory resolves the initial singularity by means of a bounce \cite{Chamseddine:2016uef}. In this context, the departures from general relativity (GR) are relevant only in a neighbourhood of the bounce. Away from the bounce, the cosmological dynamics are to an excellent approximation given by GR since the corrections to GR are suppressed by the limiting curvature scale.

The dynamics of Kantowski-Sachs (KS) space-time in limiting curvature mimetic gravity was first studied in Ref.~\cite{Chamseddine:2016ktu}, and then in Ref.~\cite{BenAchour:2017ivq} in the Hamiltonian formalism. The interest in vacuum KS space-time is motivated by the fact that in GR this space-time is isometric to the Schwarzschild black hole interior.  (For a study of spherically symmetric black hole space-times in mimetic gravity in the absence of the limiting curvature term in the action, see \cite{Gorji:2019rlm}.) In Ref.~\cite{Chamseddine:2016ktu} it was claimed that limiting curvature mimetic gravity achieves singularity resolution also in vacuum KS; it was then argued that a black-hole remnant with limiting curvature is formed at the end of gravitational collapse.  Since then, another version of limiting curvature mimetic gravity has been proposed that additionally introduces a term in the action that depends on the 3-dimensional Ricci scalar \cite{Chamseddine:2019fog}.  In this paper, following Ref.~\cite{Chamseddine:2016ktu} we study vacuum KS space-times in the limiting curvature theory proposed in Ref.~\cite{Chamseddine:2016uef}, and comment on the other limiting curvature theory proposed in Ref.~\cite{Chamseddine:2019fog} in the Discussion at the end of the paper.

The main result in this paper is to show that controlling the expansion rate alone is not enough to cure the curvature singularity of the KS space-time. We analyze the solutions of the equations of motion using dynamical system techniques. As observed already in Ref.~\cite{Chamseddine:2016ktu}, due to the limiting curvature scale the system goes through a sequence of bounces and recollapses. After each cycle, the contribution of the spatial curvature becomes increasingly important, and the duration of the cycles is monotonically decreasing. Our analysis shows that, contrary to previous claims, during this sequence of cycles there occurs a curvature singularity in finite proper time.  As the singularity is approached, the volume tends to a constant while the directional Hubble rates become arbitrarily large (one directional Hubble rate being positive and the other two negative). The expansion scalar (proportional to the mean Hubble rate) remains bounded and oscillates around zero; however, the frequency of these oscillations is monotonically increasing and eventually divergent. We show that the space-time is geodesically incomplete and the curvature invariants diverge at the space-time singularity. The oscillatory character of the solutions is reflected in the behaviour of the curvature invariants: the divergent frequency of the oscillations near the singularity contributes to the leading order terms of both the tidal forces as well as the curvature scalars $R$ and $R_{\mu\nu} R^{\mu\nu}$.

We also show that the correspondence between KS and the Schwarzschild black-hole interior does not hold in limiting curvature mimetic gravity. In GR, the Schwarzschild null event horizon corresponds, in the KS chart, to the constant time surface where the radial scale factor vanishes, and the mass is related to the scale factor in the angular directions (see Appendix~\ref{Appendix:KS_GR} for details).  However, unlike the GR case, the radial scale factor only vanishes in the infinite past in terms of proper time (at which point the angular scale factor diverges), and therefore the KS space-time in limiting curvature mimetic gravity cannot be matched to a null surface corresponding to the horizon of a black hole. This result emphasizes that it is important to be careful when extrapolating the KS/Schwarzschild interior correspondence beyond GR, as this correspondence is theory-dependent. (Such an extrapolation is the underlying assumption in a number of investigations about singularity resolution in the Schwarzschild black-hole interior in loop quantum cosmology~\cite{Ashtekar:2005qt,Modesto:2005zm}; in such cases it is essential to verify that the KS/Schwarschild interior correspondence continues to hold in this context also.)

Another reason motivating our interest in limiting curvature mimetic gravity is the existence of some connections between limiting curvature mimetic gravity and some quantum gravity models: in the cosmological sector, for flat FLRW space-times the dynamics are identical to the effective dynamics of loop quantum cosmology \cite{Bodendorfer:2017bjt, Liu:2017puc}. It has been shown that the correspondence with loop quantum cosmology does not hold beyond the perfectly homogenous and isotropic case, although the dynamics for Bianchi~I space-times is qualitatively similar in the two theories~\cite{Bodendorfer:2018ptp}, and the same transition rules for Kasner exponents across the bounce hold in LQC and limiting curvature mimetic gravity~\cite{Chamseddine:2016uef, Wilson-Ewing:2017vju, Wilson-Ewing:2018lyx} (in fact, these Kasner exponent transition rules hold quite generally for a large class of modified gravity theories where the singularity in the Bianchi~I space-time is replaced by a non-singular bounce \cite{deCesare:2019suk}). Further generalizations of mimetic gravity have also been constructed in order to reproduce the cosmological background dynamics of group field theory condensates~\cite{deCesare:2018cts}.

Note that there exist other approaches to constructing modified gravity theories with limiting curvature scales, for example in higher derivative extensions of GR, by effectively bounding curvature invariants \cite{Mukhanov:1991zn, Brandenberger:1993ef} (or even the components of the curvature tensor \cite{Yoshida:2018kwy}) by introducing suitable Lagrange multipliers and potentials. Non-singular bouncing cosmologies have been successfully realised in this framework~\cite{Mukhanov:1991zn, Brandenberger:1993ef, Yoshida:2017swb}, but black hole space-times remain singular~\cite{Yoshida:2018kwy}; this is similar to what we find here for the limiting curvature mimetic gravity theory proposed in~\cite{Chamseddine:2016uef}.

The outline of the paper is as follows: in Section~\ref{sec:mimetic}, we provide a brief review of limiting curvature mimetic gravity, then express the equations of motion for the KS space-time as an autonomous dynamical system, and study the system's phase portrait.  We find that physical solutions are divided into two classes, depending on the sign of the directional Hubble rate. For the remainder of the paper, we focus on solutions corresponding to shrinking two-spheres, which are the relevant ones for the comparison with the Schwarzschild black hole interior in GR.  (The white hole solutions with expanding two-spheres can be obtained by time-reversal.)  In Section \ref{Sec:Cycles}, we focus on the early time behaviour of the system, which is characterized by an alternating series of cosmological bounces and recollapses in the mean scale factor.  We demonstrate that the orbits come arbitrarily close to the fixed points of the dynamical system in this regime, but never actually reach them.  Rather, they asymptote to a phase-plane separatrix in the past.  We also study the behaviour of the various KS scale factors at early times, and find that the radial scale factor remains finite for all finite values of the cosmological proper time.  We also present some more technical results about the early time bounce-recollapse cycles in Appendix~\ref{sec:bounce-recollapse}, including calculations of the evolution of the spatial curvature and time elapsed during one cycle, and a proof that the radial scale factor vanishes at past infinity.   In Section~\ref{Sec:Singularity}, we determine the properties of the future attractor of the system and show that it corresponds to a curvature singularity: curvature scalars diverge, and timelike radial geodesics are incomplete and terminate in a deformationally strong singularity as defined by Ori \cite{Ori:2000fi}.  Finally, in Section~\ref{Sec:BH} we show that the KS solution in limiting curvature mimetic gravity cannot be matched to a Schwarzschild black hole at the horizon (the matching in GR is reviewed in Appendix~\ref{Appendix:KS_GR}), this shows that the general correspondence between Kantowski-Sachs and the interior of a black hole does not hold in all modified gravity theories.  We end with a discussion in Section~\ref{Sec:Conclusions}.

In this paper, we use the metric signature convention $(-+++)$.

\section{Kantowski-Sachs in mimetic gravity}
\label{sec:mimetic}

\subsection{Limiting curvature mimetic gravity}

The gravitational part of the action for mimetic gravity is
\be\label{ActionPrinciple}
S[g_{\mu\nu},\varphi,\lambda]=\frac{1}{8\pi G}\int\de^4x \sqrt{-g}\; \left[ \frac{1}{2}R-\lambda(g^{\mu\nu}\varphi_\mu\varphi_\nu+1)+f(\chi) \right]~,
\ee
where $\varphi$ is a scalar field, $\varphi_\mu\equiv\pa_\mu\varphi$, and $\chi=-\Box\varphi$. The field $\lambda$ is a Lagrange multiplier that implements the mimetic constraint $g^{\mu\nu}\varphi_\mu\varphi_\nu=-1$.  (Note that the opposite $(+---)$ convention for the signature of the metric is used in Refs.~\cite{Chamseddine:2016uef, Chamseddine:2016ktu}, as a result the definition of $\chi$ and the signs of the different terms in the action \eqref{ActionPrinciple} are adjusted accordingly.)

The vacuum field equations are then~\cite{Chamseddine:2016uef}
\be\label{FieldEquations}
G_{\mu\nu}=\tilde{T}_{\mu\nu} ~,
\ee
where $\tilde{T}_{\mu\nu}$ is an effective stress-energy tensor due to $\varphi$,
\be\label{EffectiveStressEnergy}
\tilde{T}_{\mu\nu}=2\lambda \varphi_\mu\varphi_\nu-g_{\mu\nu}(\chi f_{\chi}-f-g^{\rho\sigma}\varphi_\rho \pa_{\sigma}f_{\chi})-2\varphi_{(\mu} \pa_{\nu)} f_{\chi} ~.
\ee
The Lagrange multiplier $\lambda$ can be eliminated by solving the equation obtained by varying the action with respect to $\varphi$,
\be\label{MultiplierEquation}
\nabla^{\mu}( \pa_\mu f_\chi-2\lambda\varphi_{\mu})=0 ~.
\ee
In the synchronous gauge, the metric has the form
\be
\de s^2 = -\de t^2 + q_{ab} \de x^a \de x^b,
\ee
and denoting the determinant of the spatial metric by $q = V^2$,
\be
\chi = \frac{1}{V} \frac{\de V}{\de t},
\ee
showing that in these coordinates $\chi$ can be interpreted as a mean Hubble rate $\bar H = \chi/3$ (which, in general, will depend on time).

In cosmological space-times the integration constant in the solution of Eq.~(\ref{MultiplierEquation}) corresponds to an irrotational dust component called mimetic dark matter. Following Ref.~\cite{Chamseddine:2016ktu}, we set the integration constant to zero and focus on purely gravitational effects. The inclusion of mimetic dark matter (i.e., by choosing a non-zero value for this integration constant) lies beyond the scope of this paper and is left for future work.

For the so-called `limiting curvature' version of mimetic gravity theory studied in Refs.~\cite{Chamseddine:2016uef, Chamseddine:2016ktu}, the function $f(\chi)$ has two branches \cite{Brahma:2018dwx, deHaro:2018sqw, deCesare:2019pqj}
\begin{subequations}\label{eq:branches}
\begin{align}
f_{\rm\scriptscriptstyle B}(\chi)&=\frac{2}{3}\chi_{m}^2\left\{1+\frac{1}{2}q^2+\sqrt{1-q^2 } +q\arcsin(q)     \right\}  ~, \label{Eq:bounceBranch} \\
f_{\rm\scriptscriptstyle L}(\chi)&=\frac{2}{3}\chi_{m}^2\left\{1+\frac{1}{2}q^2-\sqrt{1-q^2}  -|q|\big(\arcsin|q| -\pi \big)     \right\} ~,\label{Eq:lateBranch}
\end{align}
\end{subequations}
where $q=\chi/\chi_{m}$, and $\chi_{m}$ denotes the maximal value that $\chi$ can reach in this theory.  Since the mean Hubble rate is bounded, it is hoped that this modified gravity theory will be free of singularities.  In the flat Friedmann-Lema\^itre-Robertson-Walker (FLRW) and the Bianchi~I space-times, the Big-Bang and Big-Crunch singularities of GR are replaced by a non-singular bounce, and the critical energy density at the bounce $\epsilon_{m} \propto \chi_{m}^2$ is related to $\chi_{m}$ \cite{Chamseddine:2016uef}.

In the flat FLRW and Bianchi~I space-times, the two branches of $f(\chi)$ correspond, respectively, to the bounce regime and to the low-curvature regime in the late universe with the transition between the two branches occurring when $\chi=\chi_{m}$ (the integration constants are fixed by the requirement that the effective stress-energy tensor be smooth at this transition between the two branches \cite{deCesare:2019pqj, deCesare:2018cts}).  Although it is expected that this theory can be extended beyond the sector of homogeneous space-times, it will be necessary to understand how the transition between the two branches of $f(\chi)$ occurs in a fully inhomogeneous space-time. In a cosmological context and for perturbative inhomogeneities, the transition between branches is determined by the background, and no discontinuities arise at the matching point in the evolution of scalar perturbations (since the sound speed is continuous, despite the second derivative of $f(\chi)$ being discontinuous).

Note also that another difficulty for mimetic gravity is that, at least in a cosmological context, linear perturbation theory always has either a ghost or a gradient instability \cite{Ramazanov:2016xhp, Firouzjahi:2017txv}.

\subsection{The Kantowski-Sachs space-time}
\label{Sec:KSgeneral}

The Kantowski-Sachs metric describes a (spatially) homogeneous but anisotropic space-time with spatial topology $\mathbb{R}\times S^2$; its line element is
\be\label{Eq:KSlineElement}
\de s^{2} = -\de t^{2} + a(t)^{2} \, \de R^{2} + b(t)^{2} R_{0}^{2} \, \de\Omega^{2} ~,
\ee
which is invariant under rotations, as well as translations in the $R$ direction. The dimensionless functions $a(t)$ and $b(t)$ are the radial and angular scale factors, respectively, while the constant $R_{0}$ has dimensions of length. The metric is also invariant under the transformations
\begin{equation}\label{eq:symmetry1}
b(t) \mapsto \gamma\, b(t)~, \quad R_{0} \mapsto \gamma^{-1} R_{0}~,
\end{equation}
and, separately,
\begin{equation}\label{eq:symmetry2}
a(t) \mapsto \gamma\, a(t)~, \quad R \mapsto \gamma^{-1} R~.
\end{equation}

It is convenient to introduce the parametrization
\be
a(t)=S(t) \, e^{\beta(t)} ~,~~ b(t)=S(t) \, e^{-\beta(t)/2}~,
\ee
where $S=(ab^2)^{1/3}$ denotes the mean scale factor. The expansion of a congruence of comoving observers is $\chi=3\dot{S}/S$, with the dot denoting differentiation with respect to $t$, and the mean Hubble rate is $\bar{H}=\chi/3$. The directional Hubble rates are
\begin{equation}\label{eq:hubble parameters}
H_{a} = \frac{1}{a} \frac{\de a}{\de t} = \frac{1}{3} \chi + u~, \quad H_{b} = \frac{1}{b} \frac{\de b}{\de t} = \frac{1}{3} \chi - \frac{1}{2} u~,
\end{equation}
where we have defined $u = {\de\beta}/{\de t}$. Finally, the 3-dimensional Ricci scalar on a constant-$t$ spatial hypersurface is
\begin{equation}\label{Eq:3DCurvatureKS}
\RicThree = \frac{2}{R_{0}^{2}b^{2}} = \frac{2e^{\beta}}{S^{2}R_{0}^{2}}~.
\end{equation}
The dynamics of the KS space-time in GR is reviewed in Appendix~\ref{Appendix:KS_GR}.

In limiting curvature mimetic gravity, the equations of motion for Kantowski-Sachs are
\begin{subequations}
\begin{align}
&\frac{1}{3}\chi^2=\epsilon +\tilde{\epsilon} ~,\label{EQFried0}\\
&\dot{\chi}=-\frac{3}{2}(\epsilon+p+\tilde{\epsilon}+\tilde{p})~,\label{EQRay0}\\
&\dot u+\chi u=\frac{\RicThree}{3} ~,\label{EQBeta0}
\end{align}
\end{subequations}
where $\epsilon$ and $p$ denote, respectively, the effective energy density and pressure due to the anisotropies and spatial curvature
\be\label{Eq:EffectiveEnergyDensityPressure}
\epsilon=\frac{3}{4}u^2-\frac{\RicThree}{2}~, \qquad
p=\frac{3}{4}u^2+\frac{\RicThree}{6} ~.
\ee
The effective stress-energy tensor \eqref{EffectiveStressEnergy} due to the mimetic scalar field also has the form of a perfect fluid, with an effective energy density and pressure given by
\be \label{EffectiveEnergyPressure}
\tilde{\epsilon}=\chi f_{\chi}-f ~, \qquad
\tilde{p}=-(\tilde{\epsilon}+ f_{\chi\chi}\dot{\chi}) ~.
\ee
For the function $f(\chi)$ given by Eqs.~\eqref{eq:branches},
\be
\tilde{\epsilon}=-\frac{\epsilon_{m}}{2}\left(1-\frac{q^2}{2}\pm\sqrt{1-q^2}\right)~,
\ee
with $q=\chi/\chi_{m}$ as above.  In this case, the equations of motion \eqref{EQFried0} and \eqref{EQRay0} become
\begin{align}
\frac{\chi^{2}}{3}= \epsilon \left( 1 - \frac{\epsilon}{\epsilon_{m}} \right)~, \quad
\dot{\chi}= -\frac{3}{2}\left( \epsilon +p \right)\left( 1 - \frac{2\epsilon}{\epsilon_{m}} \right) ~,
\end{align}
respectively.  Note that in these equations, anisotropies contribute to the effective energy density and pressure.  The limiting case $\chi_{m} \to \infty$ (in the decelerating branch $\dot{\chi}<0$) corresponds to GR, this is also attained dynamically in the regime $\epsilon \ll \epsilon_{m}$.  Finally, when the spatial curvature vanishes ($\beta\to-\infty$) these equations are those of the Bianchi~I space-time; in the mimetic gravity context this was studied in Ref.~\cite{Chamseddine:2016uef}.

\subsection{Dynamical systems formulation}
\label{dyn.sys}

The equations of motion for the Kantowski-Sachs space-time in limiting curvature mimetic gravity can be re-written as
\begin{subequations}\label{eq:mimetic dynamical system}
\begin{gather}
\dot\beta = u, \qquad \dot S = \frac{\chi S}{3}, \label{eq:betadot and Sdot} \\
\dot u = - \chi u -\frac{2}{3} \epsilon + \frac{1}{2} u^{2}, \quad \dot\chi = -\left( \frac{3}{2} u^{2} + \epsilon \right)\left( 1 - \frac{2\epsilon}{\epsilon_{m}} \right), \quad
\dot \epsilon = - \frac{2 \chi}{3} \left( \frac{3}{2} u^{2} + \epsilon \right),
\label{eq:KS dynamical system} \\
\chi^{2} = 3\epsilon \left( 1 - \frac{\epsilon}{\epsilon_{m}} \right). \label{eq:friedmann mimetic}
\end{gather}
\end{subequations}
The equation for $\dot \epsilon$ is obtained by differentiating the constraint \eqref{eq:friedmann mimetic} and using the equation of motion for $\dot \chi$ in \eqref{eq:KS dynamical system}.  The $\epsilon_{m} \to \infty$ limit of these equations reproduces the dynamical system (\ref{eq:GR dynamical system}) governing the vacuum KS space-time in GR, as expected.

Note that  $\beta$ does not appear undifferentiated in any of the equations of motion written in this form, while $S$ only contributes to the $\de S/\de t$ equation of motion.  As a result, it possible to first solve the equations \eqref{eq:KS dynamical system} which form a 3-dimensional autonomous dynamical system with the one constraint \eqref{eq:friedmann mimetic}.  The solution to this autonomous solution can then be used to obtain $S(t)$ and $\beta(t)$ from  (\ref{eq:betadot and Sdot}) via integration.

Defining a dimensionless time coordinate
\begin{equation}
T = t \sqrt{ \frac{\epsilon_{m}}{3}},
\end{equation}	
and rescaled variables by
\begin{equation}
E = \frac{\epsilon}{\epsilon_{m}}, \quad U = \frac{u}{2} \sqrt{ \frac{3}{\epsilon_{m}} }, \quad  X =  \frac{\chi}{\sqrt{3\epsilon_{m}}} ,
\end{equation}
the system of equations for the 3-dimensional autonomous dynamical system is
\begin{subequations}\label{eq:dimensionless sys}
\begin{gather}
\label{eq:ds} E^{\prime} = -2X(E+2U^{2}), \quad U^{\prime} = U^{2} - 3XU - E, \quad X^{\prime} = (E+2U^{2})(2E-1), \\ \label{eq:constraint} X^{2}=E(1-E),
\end{gather}
\end{subequations}
where a prime denotes $\de/\de T$.  Note that the constraint \eqref{eq:constraint} implies
\begin{equation}
E \in [0,1], \quad X \in [-1/2,1/2].
\end{equation}

\begin{figure}
\begin{center}
\includegraphics[width=0.8\textwidth]{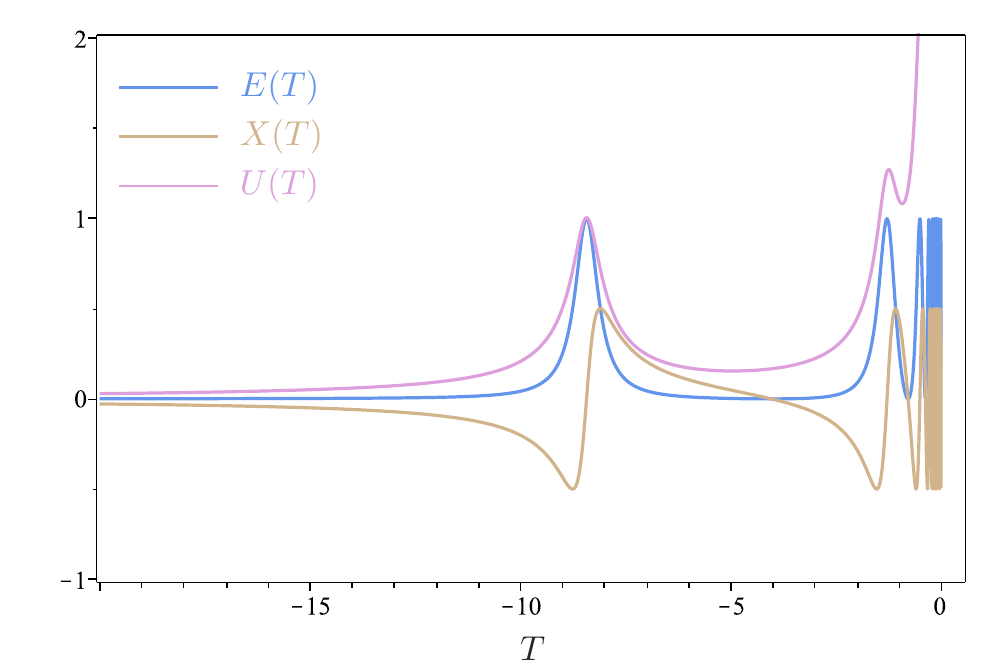}
\end{center}
\caption{Results of a numerical simulation of \eqref{eq:dimensionless sys} with initial data $X(-0.1) = E(-0.1) = 0.5$ and $U(-0.1) = 10.0$.}\label{fig:typical}
\end{figure}

Equations (\ref{eq:dimensionless sys}) can easily be solved numerically and the results of a typical simulation are shown in Figure~\ref{fig:typical}.  We observe that $U(T)$ diverges at a finite value time $T_o$ and that $X(T)$ and $E(T)$ undergo very rapid oscillations as $T \rightarrow T_o$ from the past.

To understand the origin of this behaviour, it is helpful to examine solutions of (\ref{eq:dimensionless sys}) when $U^{2} \gg 1$.  For large $U$, (\ref{eq:dimensionless sys}) can be approximated as
\begin{subequations}
\begin{gather}
E^{\prime} \approx -4XU^{2}, \quad U^{\prime} \approx U^{2}, \quad X^{\prime} \approx 2U^{2}(2E-1), \\ X^{2}=E(1-E), \quad U^{2} \gg 1.
\end{gather}
\end{subequations}
This constrained dynamical system is solvable analytically:
\begin{equation}\label{eq:large U solution}
U \approx \frac{1}{T_o - T}, \quad E \approx \frac{1}{2} + \frac{1}{2} \cos \left( \frac{4}{T-T_o} + \varphi \right), \quad X(T) \approx - \frac{1}{2} \sin \left( \frac{4}{T-T_o} + \varphi \right),
\end{equation}
where $T_o$ and $\varphi$ are constants.  Already, it is clear that (in addition to $U$ diverging as $T \to T_o$) the variables $E$ and $X$ are not well-defined for $T = T_o$; as shall be seen below $T=T_o$ corresponds to a curvature singularity in the space-time.

\begin{figure}
\begin{center}
\includegraphics[width=0.8\textwidth]{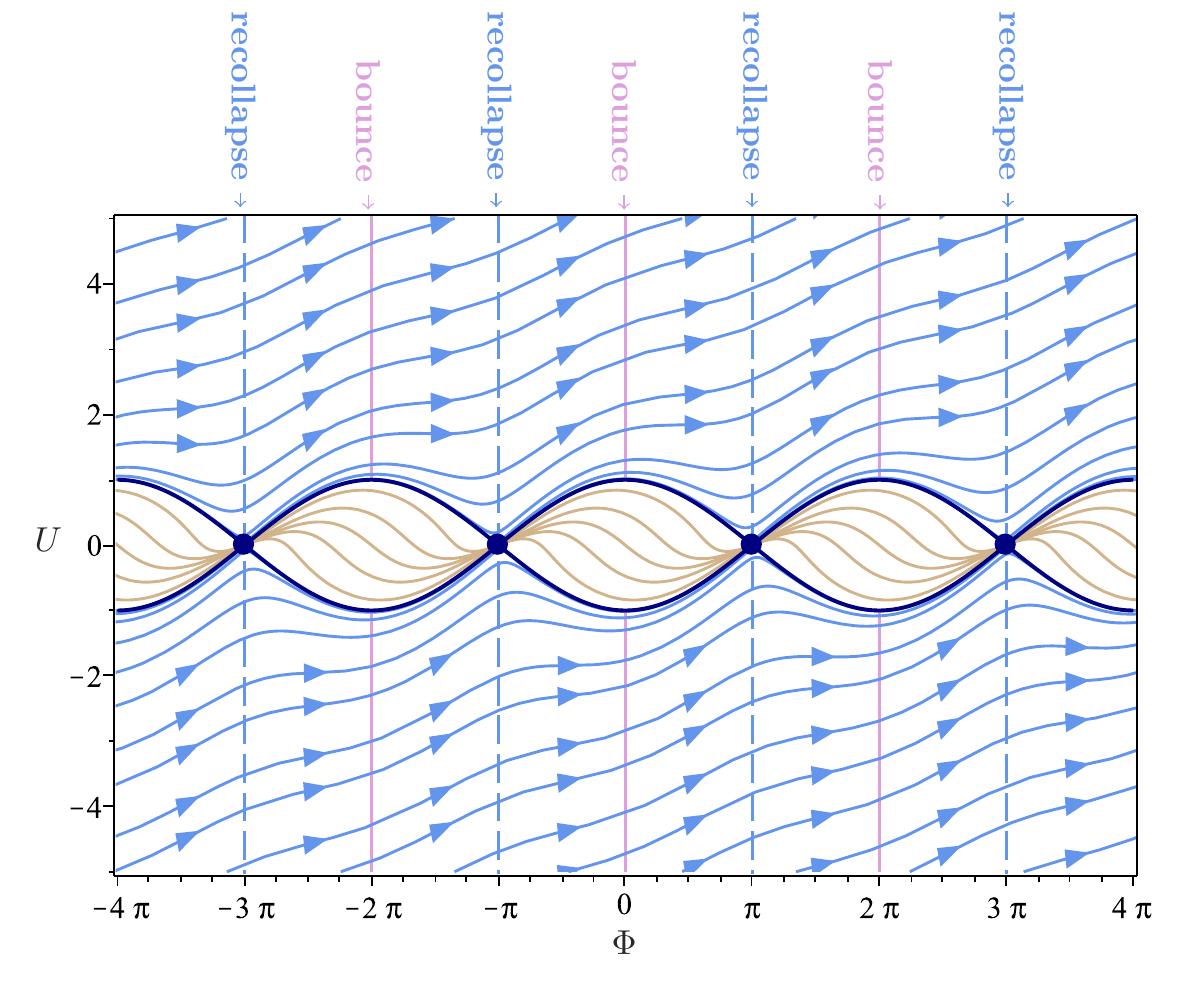}
\end{center}
\caption{
Phase portrait of \eqref{eq:ds 2D} from numeric simulations.  The blue curves have $\mathfrak{R}>0$ and the (unphysical) tan curves have $\mathfrak{R}<0$.  The fixed points are located at $(\Phi,U) = ((2n+1)\pi,0)$ with $n$ an integer, and the dark navy blue curve is the separatrix \eqref{eq:separatrix}.  Note that the trajectories flow from left to right.}\label{fig:phase}
\end{figure}

The form of the asymptotic solution \eqref{eq:large U solution} suggests the change of variables
\begin{equation}
X = \tfrac{1}{2} x, \quad E = \tfrac{1}{2} y + \tfrac{1}{2}.
\end{equation}
In terms of $x$ and $y$, the constraint becomes the equation of a circle
\begin{equation}
1 = x^{2}+y^{2},
\end{equation}
which suggests the parametrization
\begin{equation}
x(T) = \sin\Phi(T), \quad y(T) = \cos\Phi(T).
\end{equation}
This can be used to reduce the system of equations (for all $U$, not only $U \gg 1$) to that of a 2-dimensional unconstrained dynamical system:
\begin{equation}\label{eq:ds 2D}
\Phi^{\prime} = 4U^{2} + 1 + \cos\Phi, \quad U^{\prime} = U^{2} -\tfrac{1}{2} (1+\cos\Phi + 3U\sin\Phi).
\end{equation}
The phase portrait for this system from numerical simulations is shown in Figure~\ref{fig:phase}.   Defining
\begin{equation}
	\Phi_{n} \equiv n\pi,
\end{equation}
with $n$ an integer, we see that there are fixed points at $U=0$ and $\Phi = \Phi_{2n+1}$.   Note that since \eqref{eq:ds 2D} is invariant under the transformation
\be
T\mapsto -T, \quad U\mapsto -U, \quad \Phi\mapsto -\Phi,
\ee
the orbits in the lower half of the plane are obtained from those in the upper half-plane by time-reversal.  The system is also invariant under
\begin{equation}
	\Phi \mapsto \Phi + 2n\pi, \quad n \in \mathbb{Z}.
\end{equation}

The phase plane also has a separatrix.  To see this, we define the dimensionless Ricci curvature of the spatial sections of the space-time as follows:
\begin{equation}\label{eq:R def}
\mathfrak{R} \equiv \frac{{}^{(3)}\mathcal{R}}{2\epsilon_{m}} = \frac{e^{\beta}}{\epsilon_{m} R_{0}^{2}S^{2}} = U^{2}-E = U^{2} - \frac{1+\cos\Phi}{2}~.
\end{equation}
This quantity satisfies the differential equation
\begin{equation}\label{eq:R EOM 1}
\mathfrak{R}^{\prime} = 2\mathfrak{R}(U-X),
\end{equation}
which has solution
\begin{equation}\label{eq:R solution}
 \mathfrak{R}(T) = \mathfrak{R}(0) \exp \left[ 2 \int_{0}^{T} [U(\tilde T)-X(\tilde T)] \de \tilde T \right].
\end{equation}
This solution implies that the sign of $\mathfrak{R}$ is preserved along the orbits of \eqref{eq:ds 2D}.  As a result, orbits with initial data satisfying $\mathfrak{R} \ne 0$ cannot cross the $\mathfrak{R} = 0$ curve in the phase $(\Phi,U)$ plane, and this implies that the curve $\mathfrak{R} = 0$, or equivalently the curve
\begin{equation}\label{eq:separatrix}
2U^{2} - 1-\cos\Phi = 0,
\end{equation}
is a separatrix.  Note that trajectories with $\mathfrak{R}<0$ do not correspond to a KS space-time since $\mathfrak{R}<0$ would require $e^{\beta}/S^{2} < 0$, which is impossible for $\beta$ and $S$ both real.\footnote{Solutions with $\mathfrak{R}<0$ correspond to the anisotropic open universe with line-element $\de s^2 = -\de t^2 + S(t)^2 e^{2\beta(t)}\de R^2 + S(t)^2 e^{-\beta(t)}(\de \xi^2 + \sinh^2\xi\,\de\phi^2)$, which has spatial curvature $\RicThree = - {2e^{\beta}}S^{-2}R_{0}^{-2}$.}  From Figure~\ref{fig:phase}, it appears that the separatrix is a past attractor for the dynamics of trajectories with $U>0$ and $\mathfrak{R}>0$.  We will prove this in Section~\ref{Sec:Past Attractor} below.

The geometric interpretation of the dynamics of this system can be understood most directly through $x = \sin\Phi$ which is proportional to the mean Hubble rate of the space-time; clearly the mean Hubble rate vanishes for $\Phi_m = m\pi$, and there is a bounce in the mean scale factor $S$ at $\Phi_m$ for even $m$, while there is a recollapse at $\Phi_m$ for odd $m$.  Furthermore, $E$ (proportional to the effective energy density) is maximized for $\Phi_m$ with $m$ even, while it is minimized for $\Phi_m$ with $m$ odd.  The phase portrait clearly shows that the lines $\Phi=\Phi_{m}$ are intersected exactly once by the physical orbits with $\mathfrak{R}>0$. Thus, the geometry goes through an infinite sequence alternating bounces and recollapses, as shown in Figure~\ref{fig:phase}.

Using the equations of motion, the curvature invariants can be expressed as functions of the dimensionless variables $U$ and $\Phi$ only,
\begin{subequations}\label{Eq:CurvatureInvariants1}
\begin{align}
R&=4\epsilon_{m}(1+\cos\Phi)U^2~,\\
R_{\mu\nu}R^{\mu\nu}& 
= \tfrac{4}{3} \epsilon_m^2 \cos^4(\tfrac{1}{2}\Phi ) \left[ \cos^4(\tfrac{1}{2}\Phi) + 4 U^2 \cos^2(\tfrac{1}{2}\Phi) + 16 U^4\right]
~,\\
C_{\mu\nu\rho\sigma}C^{\mu\nu\rho\sigma}&=\tfrac{16}{3} \epsilon_m^2 \left[ \cos ^2 (\tfrac{1}{2} \Phi )+U\sin\Phi -2U^{2} \right]^2~,
\end{align}
\end{subequations}
while the Kretschmann invariant is related to these curvature scalars by $R_{\mu\nu\rho\sigma} R^{\mu\nu\rho\sigma} = C_{\mu\nu\rho\sigma} C^{\mu\nu\rho\sigma} + 2 R_{\mu\nu} R^{\mu\nu} - \frac{1}{3}R^2$.
Note that at the bounces $\Phi=\Phi_{2n}$, the curvature invariants are finite and are entirely determined by the spatial curvature and the limiting curvature scale $\epsilon_m$,
\begin{subequations}\label{Eq:InvariantsAtBounce}
\begin{align}
R \big|_{\Phi = \Phi_{2n}}&= 8\epsilon_{m} (1+\mathfrak{R})  ~,\\
R_{\mu\nu}R^{\mu\nu} \big|_{\Phi = \Phi_{2n}} &=\tfrac{4}{3}\epsilon_{m}^2 \left(21+36\mathfrak{R}+16 \mathfrak{R}^2\right) ~,\\
C_{\mu\nu\rho\sigma}C^{\mu\nu\rho\sigma} \big|_{\Phi = \Phi_{2n}} &=\tfrac{16}{3}\epsilon_{m}^2 \left(1+2\mathfrak{R}\right)^2 ~.
\end{align}
\end{subequations}
As shall be shown below in Section~\ref{Sec:FixedPts}, for $U>0$ the value of the curvature scalars at the bounces are bounded near the past attractor (corresponding to $\mathfrak{R}=0$).  Indeed, at the fixed points these curvature invariants vanish
\begin{equation}
	R \big|_{(\Phi,U) = (\Phi_{2n+1},0)} = R_{\mu\nu}R^{\mu\nu}  \big|_{(\Phi,U) = (\Phi_{2n+1},0)} = C_{\mu\nu\rho\sigma}C^{\mu\nu\rho\sigma} \big|_{(\Phi,U) = (\Phi_{2n+1},0)} = 0~.
\end{equation}
Conversely, as shown in Section~\ref{sec:future attractor}, these curvature invariants diverge at the future attractor, where the spatial curvature grows without bound.

\section{Early time behaviour}
\label{Sec:Cycles}

The phase plane shows two main features of interest.  First, there are a number of oscillations in the trajectory as it flows towards the right, where each complete cycle corresponds to a bounce and a recollapse in the mean Hubble rate, proportional to $\chi = \tfrac{\sqrt{3 \epsilon_m}}{2} \, \sin \Phi$.  Second, the trajectory reaches large values of $U$ at late times (for $U>0$, and the time reverse of this occurs for $U<0$).  In this section, we will study the systems at early times when the orbits are close to the separatrix.  In Section~\ref{Sec:Singularity}, we examine the large $U$ limit, which is singular.

We will only consider the $U>0$ half-plane, as the $U<0$ solutions can be obtained by time reversal.  In addition, the KS space-times have positive $\mathfrak{R}$, so we will consider solutions that lie above the uppermost separatrix, i.e., the dark navy blue curve in the upper half-plane in Figure~\ref{fig:phase}.

In appendix \ref{sec:bounce-recollapse}, we present a few additional technical results on the early time bounce and recollapse cycles, including approximate calculations of the the evolution of the spatial curvature and time elapsed during a single cycles, and a proof that the directional scale factor $a$ vanishes as $T \to -\infty$.

\subsection{Dynamics near the fixed points}
\label{Sec:FixedPts}

In this section, we consider the qualitative behaviour of the system in the vicinity of the fixed points at $(\Phi,U)=(\Phi_{2n+1},0)$.  In particular, we demonstrate that the physical trajectories with $\mathfrak{R}>0$ neither originate nor terminate at the fixed points.  Rather, the fixed points behave like repulsive saddle points for this class of trajectories: solutions come close to the fixed points but ultimately pass them by.

The usual procedure to classify a fixed point $\textbf{x}_{0}$ of an autonomous dynamical system $\mathbf{x}' = \mathbf{F}(\mathbf{x})$ is to linearize the system about the fixed point to obtain $\delta\mathbf{x}' = J\delta\mathbf{x}$, where $J$ is the Jacobian matrix of $\mathbf{F}(\mathbf{x})$ evaluated at $\mathbf{x}=\mathbf{x}_{0}$ and $\delta \mathbf{x} =  \mathbf{x}-\mathbf{x}_{0}$.  If the eigenvalues of $J$ are all non-zero, then the classification of the fixed point is straightforward.  However, this procedure fails for (\ref{eq:ds 2D}) because the Jacobian matrix is identically equal to zero at the fixed points.  We therefore have to employ different means to study the qualitative behaviour of the system.

Since (\ref{eq:ds 2D}) is invariant under $\Phi \mapsto \Phi + 2\pi$, we can without loss of generality restrict our attention to the vertical strip $\Phi \in [0,2\pi]$.  Within this sub-region of the phase plane, there is only one fixed point at $(\Phi,U)=(\pi,0)$ and we have that
\begin{equation}
	0 \le \sin(\tfrac{1}{2} \Phi) \quad \Rightarrow \quad \sin(\tfrac{1}{2}\Phi) = \sqrt{ 1 - \cos^{2}(\tfrac{1}{2}\Phi)}~.
\end{equation}
We consider a change of variables:
\begin{equation}\label{eq:change of variable}
	\eta = \tfrac{1}{2} [U+\cos(\tfrac{1}{2}\Phi)], \quad \xi = \tfrac{1}{2} [U-\cos(\tfrac{1}{2}\Phi)]~.
\end{equation}
In terms of these variables, the fixed point is at $(\xi,\eta) = (0,0)$, and the dimensionless Ricci 3-curvature is
\begin{equation}\label{eq:R-eta-xi}
	\mathfrak{R} = 4\eta\xi~.
\end{equation}
Since $\eta - \xi = \cos(\tfrac{1}{2}\Phi)$, the dynamics are restricted to the region
\begin{equation}
	-1 \le (\eta - \xi) \le 1~.
\end{equation}
Trajectories with $\mathfrak{R}>0$ and $U>0$ are further restricted to the portion of the $\xi\eta$-plane with $\xi>0$ and $\eta>0$.\footnote{For more general orbits, the lines $\xi=0$ and $\eta=0$ together are the separatrix in the $\xi\eta$-plane.} 

\begin{figure}
\begin{minipage}[t]{0.47\textwidth}
\includegraphics[width=0.95\textwidth]{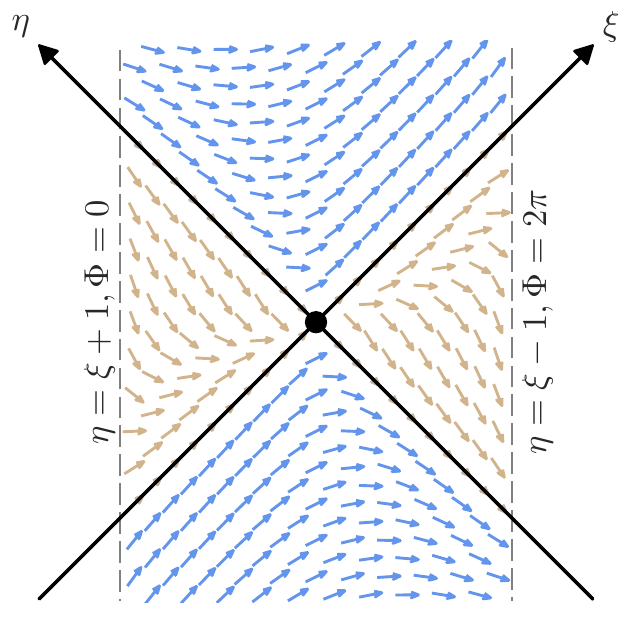}
\subcaption{Direction field plot for the dynamical system (\ref{eq:ds eta xi})}\label{fig:direction field}
\end{minipage}%
\hfill
\begin{minipage}[t]{0.47\textwidth}
\includegraphics[width=0.95\textwidth]{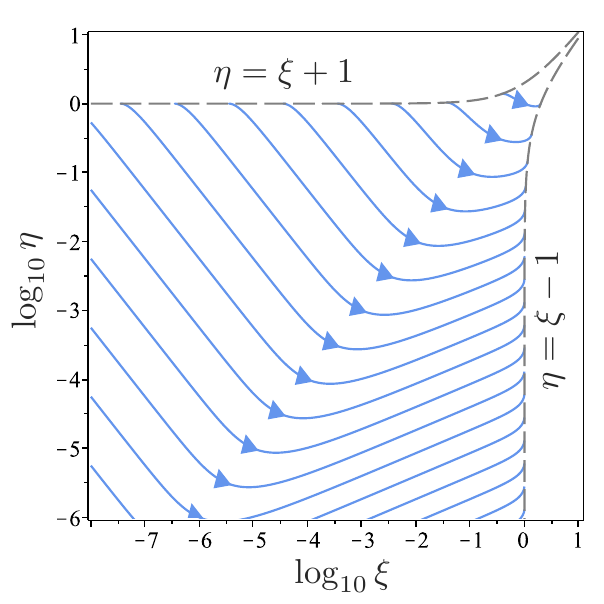}
\subcaption{Numerical solutions of  (\ref{eq:ds eta xi}) with $\eta>0$, $\xi > 0$, and $\mathfrak{R}>0$.  In this plot, the fixed point is located at $(\log_{10}\xi,\log_{10}\eta) =(-\infty,-\infty)$.}\label{fig:orbits}
\end{minipage}%
\caption{Direction field plot and numerical solutions for the dynamical system (\ref{eq:ds eta xi}).  The blue and tan arrows on the left refer to orbits with $\mathfrak{R}>0$ and $\mathfrak{R}<0$, respectively.  We see that trajectories in the first quadrant of the $\xi\eta$-plane do not terminate or end at the fixed point at $(\xi,\eta)=(0,0)$.
}\label{fig:phase_plane_direction_field}
\end{figure}

The transformed dynamical system is
\begin{equation}\label{eq:ds eta xi}
	\eta' = -\eta [ (3\eta + \xi) \sqrt{1 - (\eta-\xi)^{2}} - 2\xi], \quad \xi' = \xi [ (\eta + 3\xi) \sqrt{1 - (\eta-\xi)^{2}} + 2\eta]~.
\end{equation}
In Figure~\ref{fig:direction field}, we show a direction field plot for this dynamical system and in Figure~\ref{fig:orbits} we show numerical solutions for the orbits in the first quadrant of the $\xi\eta$-plane.  Both plots indicate that trajectories with $\eta>0$ and $\xi>0$ do not begin or terminate at the fixed point.

We can analytically confirm this conclusion by examining (\ref{eq:ds eta xi}) in the limit $|\eta - \xi| \ll 1$ (note that this includes the immediate neigbourhood of the fixed point).  To leading order, we obtain
\begin{equation}\label{eq:ds eta xi asymptiotic}
	\eta' \simeq \eta (\xi-3\eta), \quad \xi' = 3 \xi (\eta + \xi), \quad |\eta-\xi| \ll 1~.
\end{equation}
One can easily check by differentiation that the solution curves to this system are implicitly given by
\begin{equation}\label{eq:exact 1}
	0 \simeq (\xi+3\eta)^{2}k - \xi \eta^{3}~,
\end{equation}
where $k$ is a constant.  This formula implies that
\begin{equation}
	\frac{4k}{\mathfrak{R}} \simeq \left( \frac{\eta}{\xi+3\eta} \right)^{2},
\end{equation}
which in turn means that trajectories with $k>0$ must necessarily have $\mathfrak{R}>0$ and trajectories with $k<0$ must have $\mathfrak{R}<0$.  Now, equation (\ref{eq:exact 1}) can be solved for $\xi$ as a function of $\eta$:
\begin{equation}
	\xi \simeq \frac{\eta^{3}}{2k} - 3\eta \pm \frac{\eta^{2}\sqrt{\eta^{2}-12k}}{2k} ~.
\end{equation}
This implies that any trajectory with $k>0$ must have $|\eta| \ge 2\sqrt{3} k$.  Hence, all trajectories with positive spatial curvature $\mathfrak{R} > 0$ cannot approach the fixed point at $(\xi,\eta) = (0,0)$.  Conversely, non-physical trajectories with $\mathfrak{R} < 0$ approach the fixed point along the line $\xi = - 3\eta$ as $T \rightarrow \pm \infty$.

\begin{figure}
\center{
\includegraphics[width=0.7\textwidth]{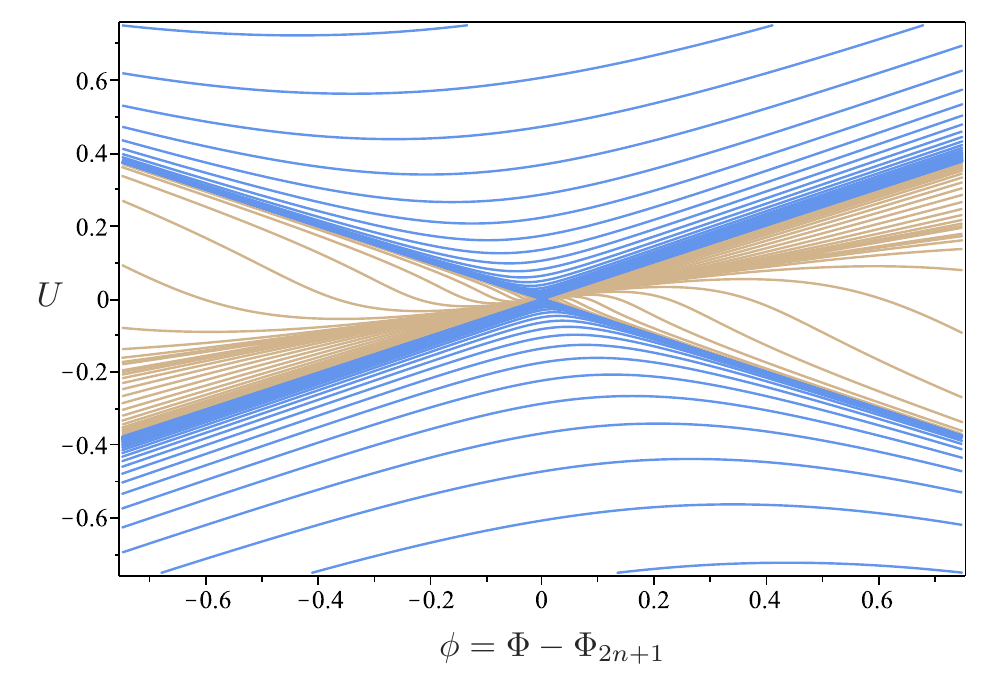}}
\caption{Phase portrait of \eqref{eq:ds 2D} in the vicinity of a fixed point $(\Phi,U)=(\Phi_{2n+1},0)$ as generated from the analytic asymptotic solution \eqref{eq:exact}.  The blue curves have $k>0$ and $\mathfrak{R}>0$, while the tan curves have $k<0$ and $\mathfrak{R}<0$. In this regime, the separatrix can be approximated by the straight lines $2U-\phi=0$ and $2U+\phi=0$, with $\phi=\Phi-\Phi_{2n+1}$. Note that trajectories flow from left to right. The left-right asymmetry in the orbits is due to the growth of the spatial curvature through the recollapse phase.}\label{fig:phase_close_up}
\end{figure}

We can rewrite the approximate solution (\ref{eq:exact 1}) in terms of the original coordinates by linearizing the transformation (\ref{eq:change of variable}) about the fixed point.  We obtain
\begin{equation}\label{eq:exact}
	0 \simeq 64k(4U-\phi)^{2} - (2U-\phi)^{3}(2U+\phi)~,
\end{equation}
with $\phi = \Phi - \pi$.  Note that due to the invariance of the dynamical system (\ref{eq:ds 2D}) under $\Phi \mapsto \Phi + 2\pi$, this relation gives the orbits of the system near all the fixed points if we identify $\phi = \Phi-\Phi_{2n+1}$.  In Figure~\ref{fig:phase_close_up}, we plot the phase portrait obtained from (\ref{eq:exact}).

Finally, we note that close to the fixed points ($U^{2} \ll 1$ and $\phi^{2} \ll 1$), the curvature invariants have the form
\begin{subequations}
\begin{align}
R&\simeq2\epsilon_{m}U^2\phi^2  ~,\\
R_{\mu\nu}R^{\mu\nu}&\simeq \tfrac{1}{192} \epsilon_{m}^2 \phi ^4  \left(256 U^4+16U^2 \phi ^2+\phi ^4\right)   ~,\\
C_{\mu\nu\rho\sigma}C^{\mu\nu\rho\sigma} &\simeq \tfrac{1}{3} \epsilon_{m}^2 \left(8 U^2+4 U \phi -\phi^2\right)^2 ~,
\end{align}
\end{subequations}
with higher-order corrections given by homogeneous polynomials of a higher degree.  At this order of approximation, $R_{\mu\nu\rho\sigma} R^{\mu\nu\rho\sigma} = C_{\mu\nu\rho\sigma} C^{\mu\nu\rho\sigma}$.

\subsection{The separatrix as a past attractor}
\label{Sec:Past Attractor}

In this section, we demonstrate that all orbits of the system with $U>0$ and $\mathfrak{R}>0$ approach the separatrix as $T \rightarrow -\infty$.  

We first note that $\mathfrak{R}>0$ implies that $\Phi'>0$ everywhere, including at the intersection of the trajectory with the vertical lines $\Phi=\Phi_{2n+1}$.  Since this holds for all $n$, $\Phi' > 0$ everywhere for trajectories with $\mathfrak{R}>0$ and as a result $\Phi \to -\infty$ as $T \to -\infty$.  

For trajectories with $U>0$ and $\mathfrak{R}>0$, from Eq.~\eqref{eq:R def}
\begin{equation}
U > \sqrt{\frac{1+\cos\Phi}{2}}~,
\end{equation}
which implies
\be \label{u-x>0}
U - X >  \sqrt{\frac{1+\cos\Phi}{2}} - \frac{\sin\Phi}{2} \ge 0 ~.
\ee
Comparison with (\ref{eq:R EOM 1}) yields that $\mathfrak{R}' \ge 0$; i.e., $\mathfrak{R}(T)$ is strictly non-decreasing for all $T$.  Now, consider the limit
\begin{equation}
	L = \lim_{T\rightarrow -\infty} \int_{0}^{T}[U(\tilde T)-X(\tilde T)]\de \tilde T ~.
\end{equation}
For this limit to be finite the integrand must tend to zero as $\tilde T \to -\infty$.  But it is clear that the limit of $U(\tilde T)-X(\tilde T)$ as $\tilde{T} \to -\infty$ does not even exist, so $L$ must be divergent, i.e.~$L \to -\infty$.  Putting this into \eqref{eq:R solution}, we obtain
\begin{equation} \label{R->0}
\lim_{T\rightarrow -\infty} \mathfrak{R}(T) = 0 ~.
\end{equation}
Since we already know from Section~\ref{Sec:FixedPts} that orbits with $\mathfrak{R}>0$ and $U>0$ do not originate at a fixed point, we conclude that the separatrix itself is a past attractor invariant manifold of the system.

Finally, this analysis shows that the dynamics of this system (for $\mathfrak{R}>0$ and $U>0$) are past complete, in the sense that given initial conditions $U(T_1)>0$ and $\Phi(T_1)$ such that $\mathfrak{R}(T_1)>0$, it is possible to evaluate $U(T)$ and $\Phi(T)$ for all finite $T < T_1$.  (On the other hand, this will not be possible for arbitrary $T > T_1$ due to a curvature singularity that occurs at some finite $T=T_o$.)

By time reversal symmetry, the above result also implies that the separatrix is a future attractor for $\mathfrak{R}>0$ in the lower half-plane $U<0$.

\subsection{Dynamics of the directional Hubble rates and scale factors}
\label{s.dyn-ab}

\begin{figure}
\begin{center}
\includegraphics[width=\textwidth]{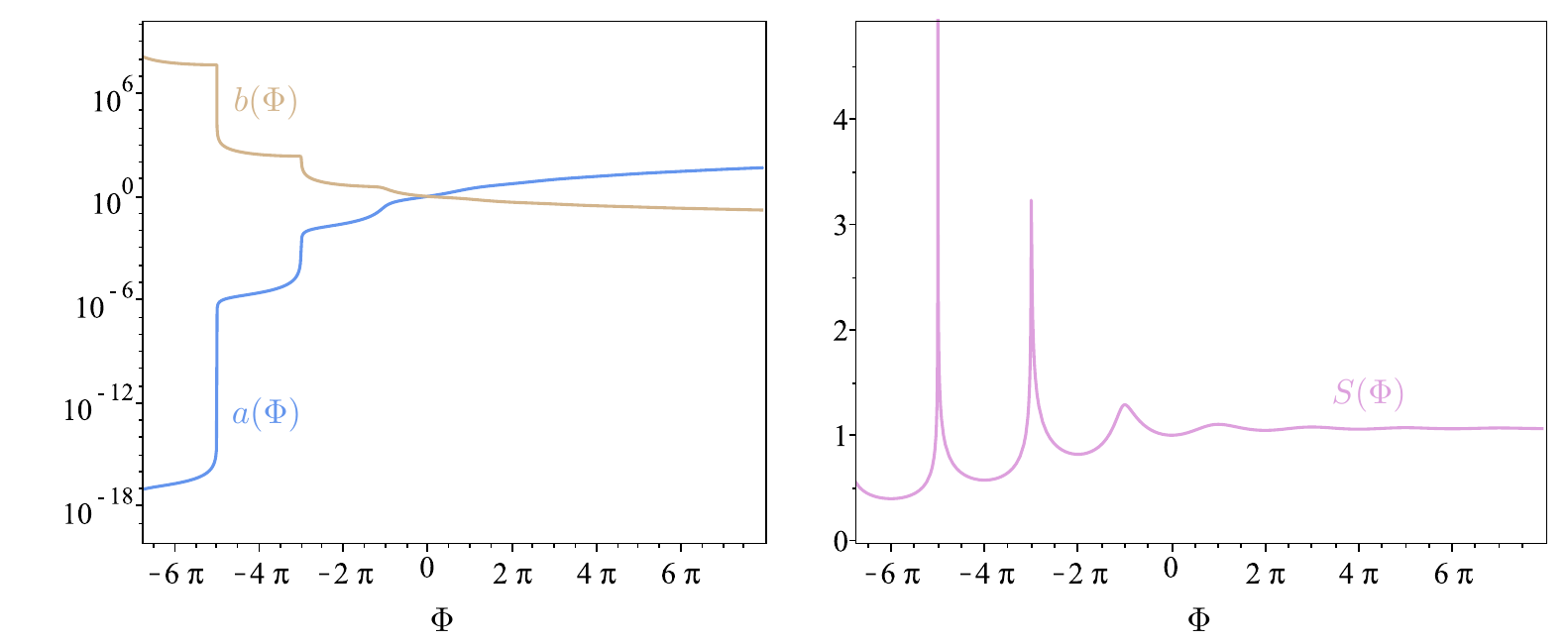}
\end{center}
\caption{Typical numerical solutions for the scale factors $a$, $b$, and $S$.  In these plots, we use $\Phi$ as a time variable.}\label{fig:scalefactor}
\end{figure}
In this section, we investigate the behaviour of the directional ($a$ and $b$) and mean ($S$) scale factors for the physical $U>0$ and $\mathfrak{R}>0$ orbits in the early time limit.  In Figure~\ref{fig:scalefactor}, we plot typical numerical solutions for these quantities for a generic choice of initial data.

We already know that $\Phi' > 0$ for trajectories with $U>0$ and $\mathfrak{R}>0$, which means that $\Phi$ is monotonic and $\chi \propto \sin\Phi$ oscillates around 0.  (Recall that $\chi$ is the mean Hubble rate.)  The oscillations in $\chi = \dot S/S$ imply in turn that the mean scale factor $S$ oscillates between expansion and contraction. The minima of $S$ occur at $\Phi=\Phi_{2n}$ and the maxima at $\Phi=\Phi_{2n-1}$.

However, the individual scale factors $a(t)$ and $b(t)$ do not oscillate.  To see this, we write the directional Hubble parameters \eqref{eq:hubble parameters} in terms of dimensionless quantities (assuming $U>0$ and $\mathfrak{R}>0$)
 \begin{equation}\label{eq:dimensionless hubble rates}
H_{a} = \sqrt{\frac{\epsilon_{m}}{3}} (2U+X)~, \qquad H_{b} = -\sqrt{\frac{\epsilon_{m}}{3}} (U-X)~.
\end{equation}
As seen in \eqref{u-x>0}, $U-X \ge 0$, and since $U >0$ it is easy to verify that $U+X \ge 0$ and $2U+X \ge 0$.  Together, these relations imply
\begin{equation}\label{Eq:HubbleRatesSigns}
H_{a} > 0~, \qquad H_{b} < 0~,
\end{equation}
for $U>0$ and $\mathfrak{R}>0$.  That is, in the $U>0$ and $\mathfrak{R}>0$ portion of the phase portrait, the scale factors $a$ and $b$ are monotonically increasing and decreasing with $T$, respectively.  This behaviour is clearly evident in the first panel of Figure~\ref{fig:scalefactor}.

Using (\ref{Eq:3DCurvatureKS}), we can also express the directional scale factors as
\be \label{def-ab}
a = R_0^2 \epsilon_m\, S^3\, \mathfrak{R}~, \quad b = \frac{1}{R_0\sqrt{\epsilon_m}} \, \mathfrak{R}^{-1/2}~.
\ee
Since we have $\mathfrak{R}\to0$ as $T\to-\infty$, it follows that $b\to+\infty$ in the same limit.  Furthermore, as demonstrated in Appendix \ref{sec:a limit}, it can be shown that
\begin{equation}
	\lim_{T\rightarrow-\infty} a(T) = 0~,
\end{equation}
and that the limit of $S(T)$ as $T\rightarrow -\infty$ does not exist.  Again, all three limits are consistent with the behaviour shown in Figure~\ref{fig:scalefactor}.

\section{The singularity}
\label{Sec:Singularity}

\subsection{Future attractor}
\label{sec:future attractor}

As reviewed in Sec.~\ref{dyn.sys}, numerical solutions indicate that (for $U>0$) after a finite time the solutions diverge, and in the limit of large $U$ it is possible to derive simple analytic (though approximate) solutions for $U, E,$ and $X$ as shown in Eq.~\eqref{eq:large U solution}.

In the large $U$ limit,
\be \label{x'}
X'(T)\simeq\frac{2}{(T-T_o)^2} \cos \left(\frac{4}{T-T_o}+\varphi\right) ~,
\ee
and the rescaled (dimensionless) mean scale factor $s = SR_{0}\sqrt{\epsilon_{m}}$ evolves as
\be \label{Eq:RescaledS}
\begin{split}
s(T) & \simeq s_o\;\exp \left\{-\frac{1}{16} (T-T_o)^2 \left[(T-T_o)\sin \left(\frac{4}{T-T_o}+\varphi \right)+2 \cos\left(\frac{4}{T-T_o}+\varphi \right)\right]\right\}\\
&\simeq s_o\left[ 1-\frac{1}{8} (T-T_o)^2 \cos\left(\frac{4}{T-T_o}+\varphi \right)-\frac{1}{16}(T-T_o)^3 \sin \left(\frac{4}{T-T_o}+\varphi \right) \right]~.
\end{split}
\ee
As $T \to T_o$, the mean scale factor approaches a constant value and the expansion $X$ always remains bounded; however, the frequency of oscillations is monotonically increasing, eventually diverging in the limit $T\to T_o$.  As a result, even though $X$ is bounded, the acceleration $X'$ oscillates with a divergent amplitude as can be seen in \eqref{x'}, and therefore does not admit a limit as $T \to T_o$.

To further study the late-time dynamics of the system (for the $U>0$ half-plane), it is helpful to compactify the $U$ direction through $h=\arctan{U}$, and then the dynamical system \eqref{eq:ds 2D} becomes
\begin{subequations}
\begin{align}
\Phi^{\prime}&=1+\cos (\Phi )+4 \tan ^2(h) ~,\label{Eq:PhiPertNearAttractor}\\
h^{\prime}&=-\tfrac{1}{2} \cos^2(h) \Big[1+\cos (\Phi )+3 \tan (h) \sin (\Phi )-2 \tan^2(h)\Big]~.\label{Eq:hPertNearAttractor}
\end{align}
\end{subequations}
It is straightforward to verify that $h=\pi/2$ is an attractor and $h = -\pi/2$ is a repeller for the system.  The velocity $\Phi^{\prime}$ is always non-vanishing away from the fixed points, and actually diverges as $h\to\pm\frac{\pi}{2}$.  Note that $h = \pm \pi/2$ are not solutions but rather belong to the boundary of the space of solutions; they are approached as $\Phi \rightarrow \infty$ or, equivalently, as $T \to T_o$.  In a neighbourhood of $h=\frac{\pi}{2}$, it is helpful to use singular perturbation theory:  we first rescale $T\to \lambda T$ and $\Phi\to \lambda^{-1}\,\Phi$, and then introduce the perturbative expansions $h=\frac{\pi}{2}+\lambda\, \delta h_1+\lambda^2\, \delta h_2 +\dots$, $\Phi=\Phi_1+\lambda\, \Phi_2 +\dots$; to first order in singular perturbation theory this gives
\be\label{Eq:FirstOrderSingPert}
\delta h_1^{\prime}\simeq1~,~~ \Phi^{\prime}_1\simeq \frac{4}{\delta h_1^2} ~,
\ee
whose solution is
\be\label{PerturbativeSolutionFutureAttractor}
\delta h_1 = T - T_o ~,  \qquad \Phi_1 = -\left(\frac{4}{T-T_o}+\varphi\right) ~.
\ee
Here $T_o$ is a constant of integration which determines the time when the space-time becomes singular.  To see this, we use the leading order perturbative result and note first that $\Phi$ diverges as $T \to T_o$, and second that
\be
h\sim\frac{\pi}{2}-\frac{4}{\Phi+\varphi} ~
\ee
approaches $\pi/2$ (which is reached at the singular time $T_o$), corresponding to an infinite $U \sim \frac{1}{4}(\Phi+\varphi)$.  As shall be shown below, this is a curvature singularity.

To second order in perturbation theory,
\be
\delta h_2^{\prime}=\frac{3}{2}\delta h_1 \sin\Phi_1 ~,~~\Phi_2^{\prime}=-\frac{8\, \delta h_2}{\delta h_1}~,
\ee
whose solution is
\begin{align}
\delta h_2&=-\frac{3}{8}(T-T_o)^3 \cos\left(\frac{4}{T-T_o}+\varphi\right)+\mathcal{O}(T-T_o)^5 ~,\\
\Phi_2&=-\frac{3}{4}(T-T_o)^2 \sin\left(\frac{4}{T-T_o}+\varphi\right)+\mathcal{O}(T-T_o)^4 ~.
\end{align}
(We do not include integration constants here, since they can be reabsorbed into $T_o$ and $\varphi$ introduced earlier.)  These results can be used to compute the asymptotics of $U$, including sub-leading order corrections to Eq.~(\ref{eq:large U solution}) which need to be taken into account in order to get the correct result for the oscillatory terms in the directional scale factors. We have
\be
h\simeq \frac{\pi}{2}+T - T_o-\frac{3}{8}(T-T_o)^3 \cos\left(\frac{4}{T-T_o}+\varphi\right) ~,
\ee
which implies
\be\label{eq:asymptoticsU}
U=\tan(h)\simeq \frac{1}{T_o-T}+(T - T_o)\left[\frac{1}{3}-\frac{3}{8}\cos\left(\frac{4}{T-T_o}+\varphi\right) \right] ~.
\ee

The evolution of the directional scale factors for large $U$ is given by
\begin{subequations}\label{eq:asymptotic scale factors}
\begin{align}\label{Eq:AsymptoticScaleFactor1}
a(T) & =\frac{s^3 (U^2-E)}{R_0\sqrt{\epsilon_m}}\simeq \frac{s_o^3}{R_0\sqrt{\epsilon_m}}\left[(T-T_o)^{-2}-\frac{1}{24}\left(28+3\cos\Phi \right) \right],
\\ \label{Eq:AsymptoticScaleFactor2}
b(T) & =\frac{1}{R_0\sqrt{\epsilon_m}\sqrt{U^2-E}}\simeq \frac{T_o -T}{ R_0\sqrt{\epsilon_m}}\left[ 1+\frac{1}{24}(T_o-T)^2\left(14-3\cos\Phi\right)\right].
\end{align}
\end{subequations}
Note that $a(T)$ diverges and $b(T)$ vanishes in the limit $T \to T_o$.  Also, even if the oscillatory terms in $a(T)$ and $b(T)$ are bounded and subdominant compared to the leading order terms as the singularity is approached, the second derivatives of these terms are nonetheless divergent and can contribute (in some cases significantly) to tidal forces in the space-time.

In addition, to leading order in $(T-T_o)$, the directional Hubble rates are
\be
H_a = \frac{2}{|T-T_o|},
\qquad \qquad
H_b = - \, \frac{1}{|T-T_o|}.
\ee
(Note that $|T-T_o|=-(T-T_o)$ for $T<T_o$.)  So although $X \propto H_a + 2H_b$ remains bounded at all times, the directional Hubble rates $H_a$ and $H_b$ both individually diverge as the singularity is approached.  From these calculations, it is clear that bounding $X$ alone is not sufficient to cure all singularities, as this does not imply that the directional Hubble rates, nor $X'$, will necessarily remain bounded.

\begin{figure}
\begin{center}
\includegraphics[width=0.8\columnwidth]{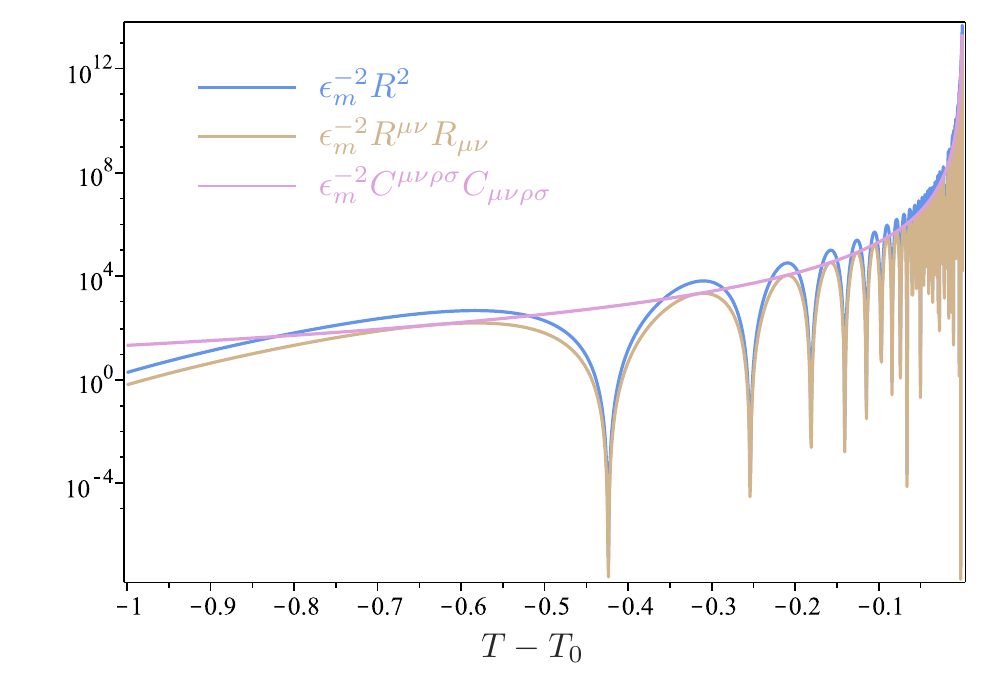}
\end{center}%
\caption{
Logarithmic plot of the various curvature invariants as the singularity at $T=T_o$ is approached.  We have assumed $\varphi=0$.\label{fig:invariants}
}
\end{figure}

Near the future attractor (for $U>0$), the curvature invariants have the form
\begin{subequations}\label{eq:late time invariants}
\begin{align}
R&\sim\frac{8 \epsilon_{m} }{(T_o-T)^2} \cos^2 \left(\frac{2}{T-T_o}+\frac{\varphi}{2}\right) ~,\\
R_{\mu\nu}R^{\mu\nu}&\sim \frac{64 \epsilon_{m}^2 }{3(T-T_o)^4}\cos^4\left(\frac{2}{T-T_o}+\frac{\varphi }{2}\right) \sim\frac{1}{3}R^2 ~,\\
C_{\mu\nu\rho\sigma}C^{\mu\nu\rho\sigma}&\sim \frac{64 \epsilon_{m}^2}{3 (T-T_o)^4}~,
\end{align}
\end{subequations}
clearly showing that there is a curvature singularity for $T = T_o$.  In particular, in terms of the original dimensionful time variable $t = T \sqrt{{3}/{\epsilon_m}}$ the Weyl scalar diverges as 
\be \label{Eq:AsymptoticsWeylMimetic}
 C_{\mu\nu\rho\sigma}C^{\mu\nu\rho\sigma}\sim 192\, (t-t_o)^{-4} ~,
\ee
where $t_o=T_o \sqrt{{3}/{\epsilon_{m}}}$.  This is similar to the late time Weyl-squared divergence that occurs in the vacuum KS space-time in GR, see Eq.~\eqref{Eq:SingularityGRasymptotics}, but in this case other curvature invariants like $R$ and $R_{\mu\nu} R^{\mu\nu}$---which automatically vanish in vacuum GR but don't in mimetic gravity---also become arbitrarily large as they oscillate with unbounded amplitude (going through 0 at the points $\Phi=(2n+1)\pi$ for all $n$) in the neighbourhood of the singular surface $T=T_o$. The late time curvature invariants (\ref{eq:late time invariants}) are plotted in Figure~\ref{fig:invariants}.

\subsection{Behaviour of geodesics near the singularity}
\label{Sec:Geodesics}

\subsubsection{Solutions to the geodesic equation}

In order to study solutions of the geodesic equation near the singularity, it is sufficient to retain only the leading order terms in the expressions for $a(T)$ and $b(T)$ in  (\ref{eq:asymptotic scale factors}). (It is a straightforward calculation to check that the next-to-leading-order terms do not affect the qualitative behaviour of geodesics near the singularity, and in particular the conclusions regarding geodesic (in)completeness.)  Note however that the oscillatory terms do contribute at leading order to the curvature scalars and the geodesic deviation equation, as shown in Sec.~\ref{ss.tidal}.

To simplify calculations in this section only, it is convenient to shift the time coordinate $t\mapsto t + t_o$ so the singularity occurs at $t=0$ and also rescale the radial coordinate by $R\mapsto S_o^{3/2} R$. Then, the asymptotic metric is
\begin{equation}
\de s^{2} \approx - \de t^{2} + \frac{\tilde{t}^4}{t^4} \, \de R^{2} + \frac{t^2}{\tilde{t}^2} \, R_{0}^{2} \de\Omega^{2}~,
\end{equation}
where $t\le 0$ and $\tilde{t}=\sqrt{3} R_0$.

Near the singularity (i.e., in the region where this asymptotic form of the metric holds), geodesics (which without any loss of generality have been assumed to lie in the $\theta=\pi/2$ equatorial plane) have two constants of the motion,
\begin{equation}
p = a^{2} \, \frac{\de R}{\de\lambda} = \frac{\tilde{t}^4}{t^4} \, \frac{\de R}{\de\lambda}~, \quad L = b^{2}R_{0}^{2} \, \frac{\de\phi}{\de\lambda} = \frac{t^2}{\tilde{t}^2} \, R_{0}^{2} \, \frac{\de\phi}{\de\lambda}~,
\end{equation}
with $\lambda$ an affine parameter.  The normalization of 4-velocities $u_{a} u^{a} = - \kappa$, with $\kappa = 1$ for time-like geodesics and $\kappa=0$ for null geodesics, yields
\begin{equation}
\kappa = \left( \frac{\de t}{\de \lambda} \right)^{2} - \frac{t^4}{\tilde{t}^4} p^{2} - \frac{\tilde{t}^2}{R_{0}^2 t^2} L^{2}~. 
\end{equation}
The affine parameter interval $\Delta\lambda$ required for a particle to travel from some initial $t = -t_{1}$ initial hypersurface in the asymptotic regime to the singularity at $t = 0$ is therefore given by
\begin{equation}
\Delta \lambda = \tilde{t}^{2} R_{0} \int_{0}^{t_{1}} \de t \frac{t}{\sqrt{R_{0}^{2}p^{2}t^{6} + \kappa R_{0}^{2} t^{2} \tilde{t}^{4} + L^{2} \tilde{t}^{6}   } }  \le \begin{cases} \frac{R_{0}t_{1}^{2}}{2\tilde{t}L}, & L \ne 0, \\ t_{1}, & L=0 \text{ and } \kappa = 1.  \end{cases}
\end{equation}
Note that $\Delta \lambda$ is infinite if $L=0$ and $\kappa = 0$, otherwise $\Delta\lambda$ is finite: radial null geodesics are complete to the future, while all other geodesics are not%
\footnote{This is analogous to the structure of a big-rip singularity in a spatially flat FLRW space-time in GR with a perfect fluid of constant equation of state $P = -\frac{4}{3}\rho$, where $P$ and $\rho$ are respectively the pressure and energy density of the perfect fluid.  See \cite{Harada:2018ikn} for a discussion of the conformal structure of big-rip cosmologies.  (The similarity is not surprising since the dominant terms in the $g_{tt}$ and $g_{rr}$ metric close to the singularity are identical.) However, there is also an important difference since in the limiting curvature mimetic KS space-time the area of the two-spheres is shrinking in the approach to the singularity, while in this limit the area of the two-spheres diverges for big-rip cosmologies. Furthermore, in the limiting curvature mimetic KS space-time $\epsilon\to0$ and $p\to+\infty$ at the singularity, which implies that the ratio $p/\epsilon$ diverges, unlike in the big-rip scenario.}.

\subsubsection{Tidal acceleration}
\label{ss.tidal}

Another way to characterize space-time singularities makes use of the acceleration between individual geodesics in a congruence.  Let $u^{\mu}$ be the 4-velocity field tangent to an affinely parametrized geodesic congruence and $q^{\mu}$ be an orthogonal spacelike unit vector:
\begin{equation}
	u^{\nu} \nabla_{\nu} u^{\mu} = 0, \quad u^{\mu}u_{\mu} = - \kappa, \quad u^{\mu} q_{\mu} = 0, \quad q^{\mu} q_{\mu} = 1,
\end{equation}
with $\kappa = 1,0,-1$ as above.  If two geodesics in the congruence are separated by a displacement $Q^{\mu} = \ell q^{\mu}$, the geodesic deviation equation gives
\begin{equation}
	\frac{1}{\ell} \frac{\text{D}^{2}Q^{\mu}}{\de \lambda^{2}} = -R_{\rho\nu\sigma}{}^{\mu}q^\nu u^\rho u^\sigma ~,
\end{equation}
where $\lambda$ is an affine parameter, as above.  If a body of finite size is travelling along a geodesic in the congruence, $\text{D}^{2} (\ell q^{\mu})/ \de \lambda^{2}$ indicates the relative tidal acceleration between parts of the body separated by a displacement $\ell q^{\mu}$. 

Let us now calculate the tidal acceleration experienced by a body flowing along an affinely parametrized timelike radial geodesic congruence in the mimetic KS space-time.  We introduce the following orthonormal basis vectors:
\begin{equation}
	t^{\mu} = (\di_{t})^{\mu}, \quad R^{\mu} = a^{-1}(\di_{R})^{\mu}, \quad \theta^{\mu} = (bR_{0})^{-1}(\di_\theta)^{\mu}, \quad  \phi^{\mu} = (bR_{0}\sin\theta)^{-1} (\di_\phi)^{\mu}~, 
\end{equation}
and define
\begin{equation}\label{eq:timelike radial congruence}
	u^{\mu} = \left( 1 + \frac{p^{2}}{a^{2}} \right)^{1/2} t^{\mu} + \frac{p}{a} R^{\mu}, \quad q_{\parallel}^{\mu} =  \frac{p}{a} t^{\mu} + \left( 1 + \frac{p^{2}}{a^{2}} \right)^{1/2} R^{\mu}, \quad q_{\perp}^{\mu} = \phi^{\mu}~.
\end{equation}
Here, $u^{\mu}$ is the 4-velocity field of the congruence satisfying $u^{\nu} \nabla_{\nu} u^{\mu}=0$ and $p$ is the conserved radial momentum.  The spacelike unit vectors $q^{\mu}_{\parallel}$ and $q^{\mu}_{\perp}$ are orthogonal to $u^{\mu}$ and characterize unit displacements parallel and perpendicular to the direction of motion, respectively. It can be checked that $u^{\mu}\nabla_\mu q^{\nu}_{\parallel}=u^{\mu}\nabla_\mu \theta^{\nu}=u^{\mu}\nabla_\mu \phi^{\nu}=0$, and therefore $\{u^{\mu},q^{\mu}_{\parallel},\theta^{\mu},\phi^{\mu}\}$ define a parallelly propagated tetrad.  The tidal acceleration induced by radial $Q^{\mu}_{\parallel} = \ell_{\parallel}q_{\parallel}^{\mu}$ or angular $Q^{\mu}_{\perp} = \ell_{\perp}q_{\perp}^{\mu}$ displacements are%
\footnote{Due to spherical symmetry, (\ref{eq:tidal forces}) is still valid if we replace $Q_{\perp}^{\mu}$ with any displacement tangent to the $R =$ constant 2-spheres and orthogonal to $u^{\mu}$.}
\begin{align}\label{eq:tidal forces}
	\frac{\text{D}^{2}Q_{\parallel}^{\mu}}{\de \lambda^{2}} = (\dot{H}_{a}+H_{a}^{2}) Q_{\parallel}^{\mu}, \quad \frac{\text{D}^{2}Q_{\perp}^{\mu}}{\de \lambda^{2}} =  \left[ \dot{H}_{b}+H_{b}^{2} +\frac{p^{2}}{a^{2}} ( \dot{H}_{b}+H_{b}^{2} - H_{a}H_{b}) \right] Q_{\perp}^{\mu}~.
\end{align}
We can relate the directional Hubble factors $H_{a}$ and $H_{b}$ to $U$ and $\Phi$ using (\ref{eq:dimensionless hubble rates}) and the formulae of Section~\ref{dyn.sys}, giving
\begin{subequations}
\begin{align}
	\frac{\text{D}^{2}Q_{\parallel}^{\mu}}{\de \lambda^{2}} = \, & \frac{1}{12} \epsilon_{m}(8U^{2}\cos\Phi + 24U^{2}-4U\sin\Phi+\cos^{2}\Phi - 2 \cos\Phi -3) Q_{\parallel}^{\mu}~, \\
	\frac{\text{D}^{2}Q_{\perp}^{\mu}}{\de \lambda^{2}}  = \, & \frac{1}{12}\epsilon_{m}(8U^{2}\cos\Phi + 2U\sin\Phi +\cos^{2}\Phi + 4\cos\Phi +3) Q_{\perp}^{\mu}\nonumber \\ 
	& + \frac{1}{6}\epsilon_{m} (1 + \cos\Phi)(4U^{2}+1+\cos\Phi) \frac{p^{2}}{a^{2}} Q_{\perp}^{\mu}~.
\end{align}
\end{subequations}
Near the singularity, we can use (\ref{eq:large U solution}), (\ref{eq:asymptotic scale factors}) and (\ref{PerturbativeSolutionFutureAttractor}) to write
\begin{equation}\label{eq:timelike acceleration}
	\frac{\text{D}^{2}Q_{\parallel}^{\mu}}{\de \lambda^{2}} \simeq \frac{2}{t^{2}} \left[ 3 + \cos\left( \frac{1}{\Omega t} +\varphi \right) \right]Q_{\parallel}^{\mu}~, \quad \frac{\text{D}^{2}Q_{\perp}^{\mu}}{\de \lambda^{2}} \simeq \frac{2}{t^{2}}  \cos\left( \frac{1}{\Omega t} +\varphi \right) Q_{\perp}^{\mu}~,
\end{equation}
we have defined $\Omega = \sqrt{3\epsilon_{m}}/12$.  We see that the parallel and perpendicular tidal accelerations are oscillatory with divergent amplitude as the singularity is approached.

We can also consider radially propagating congruences of affinely parametrized null geodesics with 4-velocity
\begin{equation}\label{eq:radial null geodesics}
u^{\mu} =  \frac{\omega_{0}}{a} ( t^{\mu} \pm R^{\mu})~, \quad u^{\nu}\nabla_{\nu} u^{\mu} = 0~,
\end{equation}
where $\omega_{0}$ is a constant corresponding to the radiation frequency when $a=1$.  Since null vectors are orthogonal to themselves, the submanifold orthogonal to $u^{\mu}$ is two dimensional with tangent space spanned by $\phi^{\mu}$ and $\theta^{\mu}$.  Hence, we only calculate the tidal acceleration due to angular displacements $Q^{\mu}_{\perp} = \ell_{\perp}q^{\mu}_{\perp}$.  Following the same steps as before, we find
\be
 \frac{\text{D}^{2}Q_{\perp}^{\mu}}{\de \lambda^{2}} = \frac{\epsilon_{m}\omega_{0}^{2}}{6a^{2}}  (1+\cos\Phi)(4U^{2}+1+\cos\Phi) Q_{\perp}^{\mu}~.
\ee
Unlike the case of timelike geodesics above, the tidal acceleration in this case vanishes as the singularity is approached---this is not surprising since the null radial geodesics are complete, unlike the time-like geodesics.

\subsubsection{The strength of the singularity}

A singularity is considered to be `strong' with respect to a given geodesic if all objects of finite size following that geodesic are either crushed or ripped apart by the singularity; otherwise, the singularity is said to be `weak' \cite{Ellis:1977pj}.

To determine whether a singularity is strong or weak, it is necessary to solve for the Jacobi fields of a given timelike geodesic with tangent vector $u^{\mu}$, given in Eq.~\eqref{eq:timelike radial congruence}.  Since the unit displacements $q_{\parallel}^{\mu}$, $q_{\perp}^{\mu}$ are parallelly propagated (see above), $\text{D}^{2}Q_{\parallel}^{\mu}/\de \lambda^{2}= \ddot{\ell}_{\parallel}q_{\parallel}^{\mu}$ and $\text{D}^{2}Q_{\perp}^{\mu}/\de \lambda^{2}= \ddot{\ell}_{\perp}q_{\perp}^{\mu}$.  For simplicity, in this section we consider only comoving geodesics with $p =0$.  Then, we obtain the following equations
\be
\frac{\ell_{\parallel}''}{\ell_{\parallel}} = \frac{a''}{a}~, \quad \frac{{\ell_{\perp}^{(1)}}''}{\ell_{\perp}^{(1)}} = \frac{b''}{b}~, \quad  \frac{{\ell_{\perp}^{(2)}}''}{\ell_{\perp}^{(2)}} = \frac{b''}{b}~,
\ee
whose general solutions are
\be\label{Eq:SolutionEllJacobi}
\ell_{\parallel}(T)=c_1\, a(T)+c_2 \, a(T)\int^{T}\frac{\de z}{a^{2}(z)}~,\quad
\ell_{\perp}^{(1,2)}(T)=c_{3,5}\, b(T)+c_{4,6} \, b(T)\int^{T}\frac{\de z}{b^{2}(z)}~,
\ee
where the $c_{i}$ are constants with the dimensions of length. The Jacobi fields define an (oriented) volume as $\mathcal{V} = \ell_{\parallel} \ell_{\perp}^{(1)} \ell_{\perp}^{(2)}$ whose behaviour at the singularity can be used to check whether the singularity is strong or not. Near the singularity, $\mathcal{V}$ has the asymptotic expansion
\be\label{Eq:VolumeExpansion}
\mathcal{V} = \frac{1}{\epsilon_m R_{0}^{2}}\left[ \, \frac{A}{T^2} + \frac{B}{T} + C + D \cos \left( \frac{4}{T} +\varphi \right)  +\mathcal{O}\left( T \right)\right] ~,
\ee
with $T = t \sqrt{\epsilon_m/3}$ and
\begin{subequations}
\begin{align}
A& = {\frac {c_{{4}}c_{{6}} \left( 1792\,\pi\,c_{{2}}+405\,c_{{1}} \right) 
}{405}}
~,\\
B& = {\frac { \left( 23\,c_{{4}}c_{{6}}\pi-3\,c_{{3}}c_{{6}}-3\,c_{{4}}c_{{
5}} \right)  \left( 1792\,\pi\,c_{{2}}+405\,c_{{1}} \right) }{1215}}
~,\\
C& = {\frac { \left( 529\,{\pi}^{2}c_{{4}}c_{{6}}-138\,\pi\,c_{{3}}c_{{6}}-
138\,\pi\,c_{{4}}c_{{5}}+36\,c_{{3}}c_{{5}}+84\,c_{{4}}c_{{6}}
 \right)  \left( 1792\,\pi\,c_{{2}}+405\,c_{{1}} \right) }{14580}}
 ~, \\
D & = \, -{\frac {c_{{4}}c_{{6}} \left( 1792\,\pi\,c_{{2}}+405\,c_{{1}}
 \right) }{1080}}
~.
\end{align}
\end{subequations}
In order to simplify these expressions, we have made use of the scaling symmetry (\ref{eq:symmetry2}) to set $s_0^3=R_0 \sqrt{\epsilon_m}$.

From the expansion~\eqref{Eq:VolumeExpansion} it is clear that, for generic values of the parameters $c_i$, the volume defined by linearly independent Jacobi fields diverges in the approach to the singularity.  Note that Tipler's definition of a strong singularity~\cite{Tipler:1977zza} is not applicable here, since it requires that $\mathcal{V}$ vanish (not diverge) at the singularity.  Of course, it is possible to extend Tipler's definition to include big-rip singularities by requiring that $\mathcal{V}$ vanish or diverge at the singularity, as suggested in Ref.~\cite{FernandezJambrina:2006hj}.  (Note however that the necessary and sufficient conditions characterizing a strong big-rip curvature singularity, as given in Ref.~\cite{FernandezJambrina:2006hj} for the conformally flat case, do not hold in anisotropic spacetimes.  This is because these conditions are obtained from those in Ref.~\cite{Clarke:1985127} by reversing the sign of $R_{\mu\nu}U^{\mu}U^{\nu}$ and only hold if the shear term in the Raychaudhuri equation is vanishing---this is not the case for anisotropic space-times.)

But even this extended definition is not sufficient.  Consider the special case $c_4=c_6=0$, which implies $A=B=D=0$ while $C\neq0$, and therefore the volume is finite in the limit $t\to0$.  This demonstrates that there are some Jacobi fields for which $\mathcal{V}$ remains nonzero and finite at the singularity, and as a result this is not a strong singularity according to the definition proposed in Refs.~\cite{Tipler:1977zza, FernandezJambrina:2006hj}.  However, this is a special and strongly fine-tuned case where two Jacobi fields diverge and the other vanishes in such a way that $\mathcal{V}$ remains finite.  While the possibility of a cancellation of this type was discussed in Ref.~\cite{Tipler:1977zza}, it was ultimately dismissed as unlikely and ignored for the initial proposal of a definition for a strong singularity.  However, Ori proposed an extended definition for a strong singularity to account for this possibility \cite{Ori:2000fi}, with a deformationally strong singularity being one where, for all possible Jacobi fields either: (i) $\mathcal{V}$ goes to zero at the singularity, or (ii) the norm of at least one Jacobi field diverges at the singularity (with the norm evaluated with respect to a set of tetrads parallelly propagated by the Jacobi fields).  As can be easily checked, in the limit $t\to0$ there exists a Jacobi field $Q^{\mu}_{\parallel}$ with an unbounded component in the parallelly propagated tetrad frame specified in \eqref{eq:timelike radial congruence}; this is true for all solutions for $Q^{\mu}_{\parallel}$ such that $1792\,\pi\,c_{{2}}+405\,c_{{1}} \neq0$ (and for the case of $1792\,\pi\,c_{{2}}+405\,c_{{1}} = 0$, $\mathcal{V}$ necessarily vanishes at the singularity satisfying condition (i)).  The definition of a deformationally strong singularity is able to capture cases where strong tidal deformations (as represented by the divergence of at least one Jacobi field) would destroy any physical object hitting the singularity, even though its volume may remain finite.  This extended definition is required in this setting.  Another interesting case arises if $c_1$ and $c_2$ satisfy $1792\,\pi\,c_{{2}}+405\,c_{{1}}=0$.  In this case, $A=B=C=D=0$ in Eq.~\eqref{Eq:VolumeExpansion}, independently of the values of $c_4$ and $c_6$, and $\mathcal{V}\to0$ as $t\to0$, giving a crushing singularity.

So for a generic choice of Jacobi fields, the result is a ripping singularity where $\mathcal{V}$ diverges at the singularity, but it is possible for certain choices of $c_i$ in the Jacobi fields for the singularity to be a deformational, or even a crushing singularity.  Therefore, this is a deformationally strong singularity since each of these three possibilities satisfies one of the two sufficient conditions according to Ori's definition \cite{Ori:2000fi}.

Note that the well-known necessary and sufficient conditions given in Ref.~\cite{Clarke:1985127} for the occurrence of a strong singularity in the sense of Tipler cannot be used in this context. This is because the derivation of these conditions requires the causal convergence condition $R_{\mu\nu}U^{\mu}U^{\nu}\geq 0$ (here $U^{\mu}$ is either timelike or null), but this condition does not always hold in limiting curvature mimetic gravity. This is due to the fact that the effective stress-energy tensor \eqref{EffectiveStressEnergy} violates both the strong and the null energy conditions in a neighbourhood of each bounce.

Finally, to avoid any confusion we stress that the curvature singularity we found is not caused by caustics of mimetic dark matter (recall that we are considering the vacuum case here with no contribution from mimetic dark matter since that the constant of integration for Eq.~\eqref{MultiplierEquation} has been set to zero).  It would be interesting to see whether or not including mimetic dark matter could generate additional singularities due to caustics as studied in, e.g., Refs.~\cite{Mukohyama:2009tp, Babichev:2016jzg}.

\section{Relation to the Schwarzschild interior}
\label{Sec:BH}

One important reason the Kantowski-Sachs space-time is of considerable physical interest is due to the isomorphism in GR between the KS space-time and the Schwarzschild interior, see Appendix~\ref{Appendix:KS_GR} for details.  In GR, there are two coordinate singularities in the vacuum KS space-time.  One corresponds to the physical singularity at the center of the Schwarzschild black hole, and the other corresponds to the surface that, from the perspective of the Schwarzschild space-time it can be embedded in, is the event horizon where the vanishing scale factor indicates that the constant $r=2GM$ surface becomes null.  (In KS coordinates, this surface corresponds to $t = 0$, see appendix \ref{Appendix:KS_GR}.)

On this surface, $a(t=0) = 0$ and $b(t=0) = 1$, while the space-time curvature remains finite, for example the Kretschmann scalar is $R^{\mu\nu\rho\sigma} R_{\mu\nu\rho\sigma} = 3/(4 G^4 M^4)$.  Therefore, given a KS space-time in GR, it can be interpreted as the interior of a Schwarzschild black hole of mass $M$, with $M$ determined by the value of the Kretschmann scalar on the constant $t$ surface corresponding to the coordinate singularity.

This naturally leads to the question of whether the KS space-time may also capture the relevant physics of a black hole interior for modified gravity theories, or perhaps for quantum gravity.  The answer to this question may depend on the specific theory of interest, but for limiting curvature mimetic gravity the KS space-time cannot be matched at the horizon to a Schwarzschild-like black hole solution.

To see this, consider $U>0$ solutions (so $a(t)$ is increasing and $b(t)$ is decreasing as inside a Schwarzschild black hole solution) of limiting curvature mimetic gravity.  But as shown in Sec.~\ref{s.dyn-ab}, the scale factor $a(t)$ is always positive for finite $t$, and approaches zero only asymptotically as $t \to -\infty$, where $b(t) \to \infty$.  Clearly, this cannot be matched to a Schwarzschild solution, which would require that at finite $t$, $a(t)=0$ and $b(t)$ has a finite value.  (Note that while this condition is not sufficient for the KS space-time to be interpreted as the interior of a black hole, it is necessary%
\footnote{The argument that the mimetic KS space-time cannot be matched onto an exterior black hole solution can also be stated in a coordinate independent way:  the KS space-time has a Killing horizon at any hypersurface where $\xi^{a} = (\di_{R})^{a}$ becomes null; i.e., on any hypersurface where $a=0$.  In the mimetic KS solution, the Killing horizon occurs in the infinite past as measured by the proper time along comoving geodesics.  Therefore, the space-time is not extendable across the Killing horizon and cannot be matched to an exterior black hole solution that possesses its own Killing horizon.}.)
Therefore, in the context of limiting curvature mimetic gravity, the KS space-time cannot be interpreted as the interior of a black hole.

As an aside, note that another approach was proposed in Ref.~\cite{Chamseddine:2016ktu} where GR is used near the horizon, and then one `switches' to the mimetic theory on an appropriate matching hypersurface inside the black-hole interior. To justify this procedure, the authors of Ref.~\cite{Chamseddine:2016ktu} argued that, at least for large black holes, the modifications to GR due to the limiting curvature mimetic gravity action only become relevant deep inside the black hole.  Requiring continuous scale factors and their first derivatives---in analogy to the Israel junction conditions---implies that the solutions of GR and limiting curvature mimetic gravity can only be matched on the hypersurface where $S$ is maximal in GR (and $S$ is locally maximal in the limiting curvature mimetic theory).  This requirement arises since continuity in the scale factors and their first derivatives from a GR solution (where $R_{\mu\nu}=0$ everywhere) is only possible at the surface where $R_{\mu\nu}=0$ in the mimetic theory also, which singles out the maximal $S$ hypersurface.  This procedure then allows a matching between a near-horizon solution satisfying the Einstein equations, and a high-curvature oscillatory solution satisfying the mimetic equations of motion that eventually reaches the singularity described in Sec.~\ref{Sec:Singularity}.  There are however a few drawbacks to this approach.  First, while on the one hand matching conditions are normally used to handle discontinuities in the matter content of a space-time, from a mathematical perspective it is equally possible to use matching conditions to change the theories determining the equations of motion governing the dynamics.  But on the other hand, some justification is needed to explain why such `hybrid' space-times should represent a valid approximation to full black-hole solutions in limiting curvature mimetic gravity. Second, if initial conditions are instead imposed in the high-curvature regime where the dynamics are governed by the limiting curvature mimetic theory and one evolves backwards in time, then there is a countably infinite number of surfaces where $R_{\mu\nu}=0$ where the mimetic solution could be matched to the GR solution, with each choice corresponding to a black hole with a different mass---therefore, the backwards evolution depends very strongly on which surface is chosen to perform the matching on, which appears to require an additional ad hoc input.

This discussion is also cautionary for the study of quantum gravity effects in the black hole interior.  Due to the classical isomorphism between the KS space-time and the interior of a Schwarzschild black hole, quantum gravity effects are often studied in KS space-times with the hope that the results will be applicable to black hole space-times.  But as this discussion shows, the relation between the KS space-time and the Schwarzschild black hole interior depends in an essential way on the dynamics, and this relation may no longer hold once quantum gravity effects are included.

\section{Conclusions}
\label{Sec:Conclusions}

The main result of this paper is that the limiting curvature mimetic gravity theory proposed in Ref.~\cite{Chamseddine:2016uef} does not cure all of the spacetime singularities of GR, and specifically the vacuum Kantowski-Sachs space-time is singular.

In this theory, due to the interplay of the limiting curvature scale and the spatial curvature the dynamics of the KS space-time consists of an infinite sequence of cycles alternating phases of expansion and contraction in the mean scale factor $S$.  While in GR the contraction phase inevitably terminates at a curvature singularity, in this case the existence of a limiting curvature scale significantly modifies the dynamics.  Instead, there is a smooth transition (bounce) connecting the contracting phase to the expanding one and then the spatial curvature is responsible for another recollapse, causing the evolution to proceed in an oscillatory fashion. Since the duration of these oscillations is monotonically decreasing and approaches zero within a finite proper time interval, there occurs a singularity at the time when the frequency diverges.  On this singular surface, curvature invariants and tidal forces for time-like geodesics either diverge or oscillate with a divergent amplitude.

More generally, this result shows that providing an upper bound to the mean Hubble rate (which is exactly what the limiting curvature mimetic theory of Ref.~\cite{Chamseddine:2016uef} does) is not sufficient to ensure that all singularities are resolved.  In particular, it is possible for individual directional Hubble rates to diverge, or for the time derivative of the mean Hubble rate to diverge, even if the mean Hubble rate remains bounded.   Either of those two possibilities necessarily leads to a singularity---in the KS space-time, both of these possibilities occur.

The second main result is that, unlike in GR, the KS solutions in limiting curvature mimetic gravity cannot be matched to a Schwarzschild black hole at the horizon. This is a cautionary example about the potential risks of extrapolating the isometry of KS spacetime with the interior of a static spherically symmetric black hole beyond general relativity.

Finally, it is important to emphasize that in this paper we only considered one particular version of a limiting curvature mimetic gravity theory, specifically the one proposed in Ref.~\cite{Chamseddine:2016uef}.  Since that paper, there have also been other proposals put forward for different limiting curvature theories within the mimetic gravity family, in particular by adding a contribution to the action which depends on the 3-dimensional curvature of the space-time \cite{Chamseddine:2019fog}.  Of course, the dynamics will be different in such a theory, and there are indications that black hole space-times in this particular theory may be non-singular \cite{Chamseddine:2019fog}, although a detailed analysis to verify this remains to be done.

\appendix

\numberwithin{equation}{section}

\section{The Kantowski-Sachs space-time and the Schwarzschild interior}
\label{Appendix:KS_GR}

\subsection{Vacuum Kantowski-Sachs in general relativity}

\begin{figure}
\begin{center}
\includegraphics[width=0.8\textwidth]{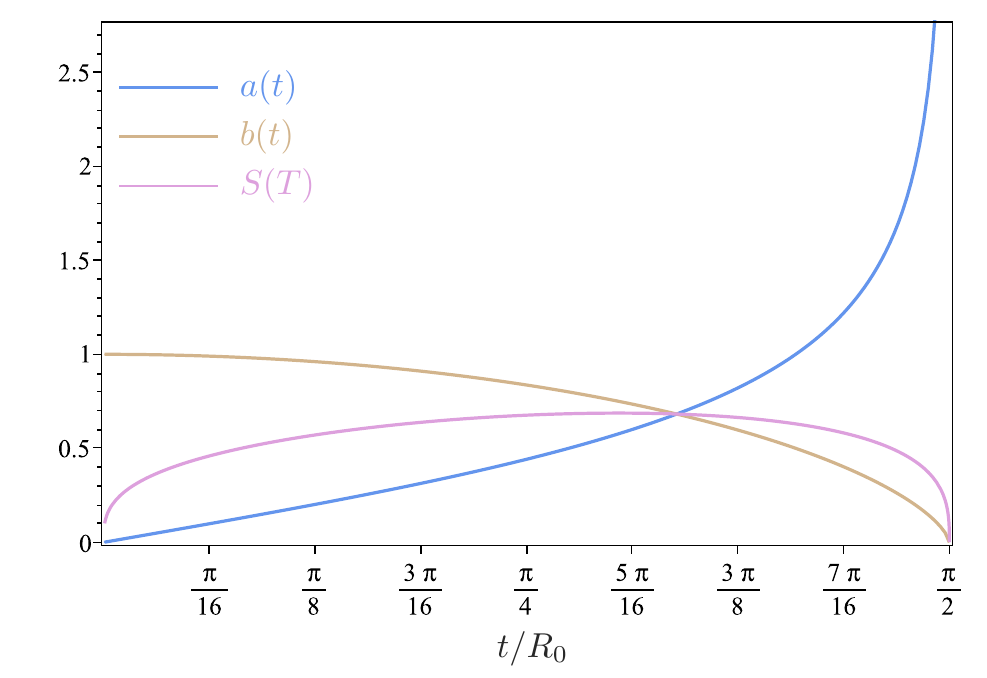}
\end{center}
\caption{Behaviour of the scale factors for the Kantowski-Sachs metric corresponding to the Schwarzschild interior.  We have selected $\sigma=-1$, which implies that $b = r/2GM$ is a monotonically decreasing function and therefore these coordinates cover the whole black-hole interior.}\label{fig:GR scale factors}
\end{figure}

The KS line element is given in Eq.~\eqref{Eq:KSlineElement}.  Assuming $a,\,b \ne 0$, the vacuum Einstein field equations reduce to
\begin{align}\label{eq:original DS}
	\ddot{a} = - \frac{2\dot{a}\dot{b}}{b}~, \quad
	\ddot{b}= - \frac{\dot{a}b}{2a}~, \quad
	\ddot{b}= -\dot{b} \left( \frac{\dot{a}}{a} + \frac{\dot{b}}{b} \right) - \frac{1}{bR_{0}^{2}}~,
\end{align}
and these can be combined to find an ODE for $b$,
\begin{equation}
2R_{0}^{2}b\ddot{b} + R_{0}^{2} (\dot{b})^{2} + 1 = 0~.
\end{equation}
This equation can be solved implicitly for $b(t)$, with the result
\begin{equation}
	t = -\sigma \frac{R_{0}}{c_{1}} \left[ \sqrt{c_{1}b(1-c_{1}b)} - \frac{1}{2} \arcsin \left( 2c_{1}b-1 \right) + c_{2} \right],
\end{equation}
where $c_{1}$ and $c_{2}$ are constants of integration and $\sigma=\pm1$.  Without any loss of generality, the constant $c_1$ can be reabsorbed using the symmetry \eqref{eq:symmetry1} with $\gamma = c_{1}^{-1}$,
\begin{equation}\label{eq:b soln}
	t = -\sigma R_{0} \left[ \sqrt{b(1-b)} - \frac{1}{2} \arcsin \left( 2b-1 \right) + c_{2} \right].
\end{equation}
This equation implies
\begin{equation}\label{Eq:bDotKS}
	\dot{b} =  \frac{\sigma}{R_{0}} \sqrt{ \frac{1}{b} - 1 }~,
\end{equation}
and combined with \eqref{eq:original DS}, this gives
\begin{equation}\label{eq:a soln}
	\frac{\de a}{\de b} = - \frac{a}{2b(1-b)} \quad \Rightarrow \quad a = a_{0} \sqrt{\frac{1}{b} - 1}~,
\end{equation}
where $a_{0}$ is an integration constant.  Note that $a_{0}$ can be reabsorbed via a coordinate transformation $R \mapsto  R/a_{0}$.  Equations \eqref{eq:b soln} and  \eqref{eq:a soln} give the most general vacuum solution for the Kantowski-Sachs space-time with both $a$ and $b$ nonzero.  These scale factors are plotted in Figure~\ref{fig:GR scale factors}.

It is possible to rewrite \eqref{eq:original DS} in terms of an autonomous dynamical system with one constraint:
\begin{subequations}\label{eq:GR dynamical system}
\begin{gather}
	\frac{\de S}{\de t} = \frac{\chi S}{3}~, \quad \frac{\de\beta}{\de t} = u~, \quad \frac{\de\chi}{\de t} = -\left( \frac{3}{2} u^{2} + \epsilon \right) ~, \quad \frac{\de u}{\de t} = -\chi u + \frac{1}{2} u^{2} - \frac{2}{3} \epsilon~, \\
	\chi^{2} = 3\epsilon~, \label{eq:friedmann GR}
\end{gather}
\end{subequations}
where the effective energy density $\epsilon$ is defined as in Eq.~\eqref{Eq:EffectiveEnergyDensityPressure}.  Note that these equations are identical to the $\epsilon_{m} \to \infty$ limit of the dynamical system (\ref{eq:mimetic dynamical system}) governing the mimetic KS space-time.

The vacuum solution has coordinate singularities at $(a,b) = (0,1)$ and $(a,b) = (\infty,0)$. Both are caustics, since the volume element $\sqrt{-g}=4\pi^2 R_0^2 S^3$ has two zeros, at $b=0$ and $b=1$, as can be seen by expressing the mean scale factor $S$ in terms of $b$ using the above results:
\begin{equation}\label{Eq:S6KS}
	S^{6} = b^{3}(1-b)~.
\end{equation}
However, the Kretschmann scalar shows that there is a genuine curvature singularity only at $b=0$:
\begin{equation}\label{Eq:KretschmannKSGR}
	R_{\mu\nu\rho\sigma} R^{\mu\nu\rho\sigma} = C_{\mu\nu\rho\sigma} C^{\mu\nu\rho\sigma} = \frac{12}{R_{0}^{4}\, b^{6}}~.
\end{equation}
At $b=1$, the Kretschmann scalar remains finite.

\subsection{Interior Schwarzschild as a Kantowski-Sachs space-time}

The Schwarzschild metric, in the interior region where $r < 2GM$, is
\begin{equation}
	\de s^{2} = \left(\frac{2GM}{r} -1 \right) \de\tau^{2} -   \left(\frac{2GM}{r} -1 \right)^{-1}\de r^{2} + r^{2} \de\Omega^{2}~.
\end{equation}
In this region $\tau$ is a spacelike coordinate, while $r$ is timelike. Consider a coordinate transformation such that
\begin{equation}\label{Eq:CoordinateTransformationSchwarzschildKS}
	\de t = \sigma \left(\frac{2GM}{r} -1 \right)^{-1/2}\de r, \quad \tau = R, \quad \sigma= \pm 1~,
\end{equation}
with the finite form of \eqref{Eq:CoordinateTransformationSchwarzschildKS} given by
\be \label{Eq:TimeSchwarzschildInterior}
t(r)= 2GM \Bigg[ \sqrt{\frac{r}{2GM}\left(1-\frac{r}{2GM}\right)}+\arccos\sqrt{\frac{r}{2GM}}\; \Bigg] ~,
\ee
having selected an integration constant so that $t=0$ when $r = 2GM$.

In these coordinates, the line element takes the Kantowski-Sachs form
\begin{equation}
	\de s^{2} = -\de t^{2} +  \left[\frac{2GM}{r(t)} -1 \right] \de R^{2} + r^{2}(t)\,\de\Omega^{2}~,
\end{equation}
and comparison with Eq.~\eqref{Eq:KSlineElement} gives
\begin{equation}\label{eq:interior 1}
	b(t) = \frac{r(t)}{2GM}~, \quad a(t) = \left[\frac{2GM}{r(t)} -1 \right]^{1/2} = \left[\frac{1}{b(t)} -1 \right]^{1/2}~, \quad R_{0} = 2GM~.
\end{equation}
These equations~\eqref{eq:interior 1} implicitly define the directional scale factors $a$ and $b$ as functions of the Kantowski-Sachs time coordinate $t$ through the Schwarzschild radial coordinate $r$.

We now derive the asymptotics of the directional scale factors  $a$ and $b$ near the zeros of $S$, using Eqs.~\eqref{Eq:TimeSchwarzschildInterior} and~\eqref{eq:interior 1}.
Near the horizon, $r\lesssim 2GM$ and
\begin{align}
a(t)&\simeq \frac{t}{4GM} ~,\\
b(t)&\simeq 1 ~.
\end{align}
And near the singularity at  $t_o=\pi\, GM$,
\begin{align}
a(t)&\simeq \left[\frac{3}{4}\left(\frac{t_o-t}{GM}\right)\right]^{-1/3} ~,\\
b(t)&\simeq \left[\frac{3}{4}\left(\frac{t_o-t}{GM}\right)\right]^{2/3}\simeq \left(a(t)\right)^{-2} ~.\label{Eq:NearSingularityAsymptoticsB}
\end{align}
Still near the singularity, using Eq.~\eqref{Eq:KretschmannKSGR} and the asymptotic formula \eqref{Eq:NearSingularityAsymptoticsB},
\be\label{Eq:SingularityGRasymptotics}
R_{\mu\nu\rho\sigma}R^{\mu\nu\rho\sigma}=C_{\mu\nu\rho\sigma}C^{\mu\nu\rho\sigma}=\frac{48 (GM)^2}{r^6}\simeq\frac{64}{27}(t_o-t)^{-4}~,
\ee
showing that the Weyl curvature diverges at the singularity.

\section{Early time bounce-recollapse cycles}
\label{sec:bounce-recollapse}

In this appendix, we present a few technical results about the early time behaviour of limiting curvature mimetic KS cosmologies in the early time limit as they undergo an infinite number of bounce and recollapse cycles in the mean scale factor.

\subsection{Change in spatial curvature during early time bounce-recollapse cycles}

\begin{figure}
\begin{center}
\includegraphics[width=0.8\textwidth]{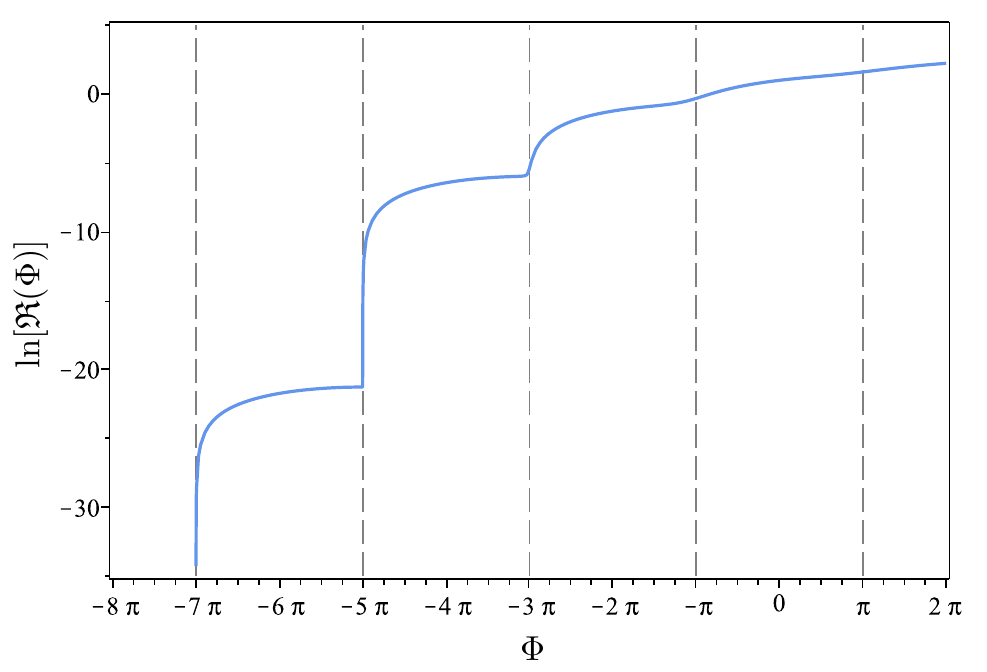}
\end{center}
\caption{Numerical solution for $\ln\mathfrak{R}$ from equation (\ref{eq:R-Phi ODE}) assuming initial data $\ln\mathfrak{R} = 1$ when $\Phi = 0$.  The dashed lines indicate recollapse times $\Phi = \Phi_{2n+1}$.  We see that for $\mathfrak{R} \ll 1$, the derivative $\de (\ln\mathfrak{R})/ \de \Phi$ is large at the recollapse times.}\label{fig:R numerics}
\end{figure}

The spatial curvature changes from one bounce-recollapse cycle to another.  Considering the early time limit $\mathfrak{R} \ll 1$ (as usual, for solutions with $\mathfrak{R} > 0$ and $U>0$), we begin by noting again that if $U \ne 0$, then (\ref{eq:ds 2D}) implies $\Phi'>0$ and we can therefore use $\Phi$ as a time coordinate.  Using both (\ref{eq:ds 2D}) and the definition (\ref{eq:R def}), we can derive the differential equation satisfied by $\mathfrak{R} = \mathfrak{R}(\Phi)$:
\begin{equation}\label{eq:R-Phi ODE} 
\frac{1}{\mathfrak{R}} \frac{\de\mathfrak{R}}{\de\Phi}= \frac{2\sqrt{\mathfrak{R}+\cos^{2}(\tfrac{1}{2}\Phi)} -\sin(\Phi)}{4\mathfrak{R}+6\cos^{2}(\tfrac{1}{2}\Phi)}~.
\end{equation}
In Figure~\ref{fig:R numerics}, we show a typical numerical solution of this equation.  We see that when $\mathfrak{R} \ll 1$, the derivative of $\ln \mathfrak{R}$ with respect to $\Phi$ becomes large to the right of the recollapse points $\Phi = \Phi_{2n+1}$.  Conversely, the derivative is small as recollapse points are approached from the left.  Both of these behaviours can be easily understood from the expansion of the righthand side of (\ref{eq:R-Phi ODE}) for $\mathfrak{R} \ll 1$ and $\Phi \sim \Phi_{2n+1}$.

\begin{figure}
\begin{center}
\includegraphics[width=0.8\textwidth]{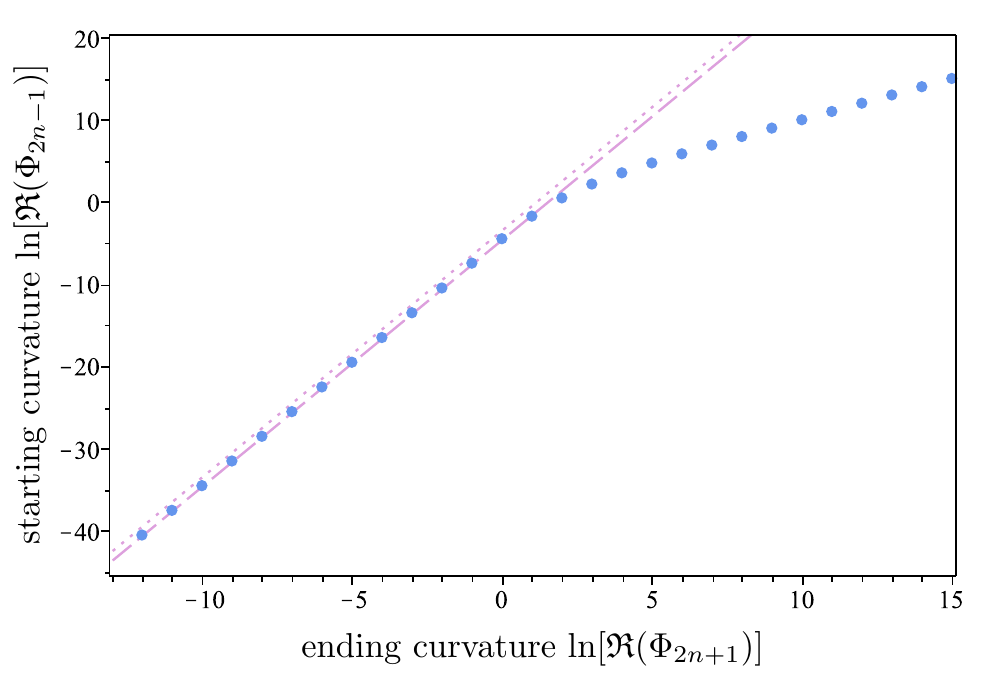}
\end{center}
\caption{Value of the spatial curvature at the beginning of a bounce-recollapse cycle as a function of its value at the end of the cycle as calculated from the numeric solutions of (\ref{eq:ds 2D}) (solid circles).  The cycle starts with a recollapse at $\Phi = \Phi_{2n-1}$ and ends at another recollapse at $\Phi=\Phi_{2n+1}$.  The dashed line shows the best fit (\ref{eq:R fit}) to the simulation data for cycles where the curvature is small; i.e.\ $\mathfrak{R}(\Phi_{2n+1}) \ll 1$. The dotted line shows the analytic approximation (\ref{eq:R analytic}).}\label{fig:R loglog}
\end{figure}
From the numerical solution of (\ref{eq:R-Phi ODE}), we can directly compute the spatial curvature at a given recollapse (corresponding to $\Phi = \Phi_{2n-1}$) as a function of its value at the next recollapse (corresponding to $\Phi = \Phi_{2n+1}$).  The results of such a calculation are shown in Figure~\ref{fig:R loglog}.  When $\mathfrak{R}(\Phi_{2n+1}) \ll 1$, the relationship between the spatial curvature at successive recollapses is well approximated by the power law
\begin{equation}\label{eq:R fit}
	\mathfrak{R}(\Phi_{2n-1}) = 0.01122 \times [\mathfrak{R}(\Phi_{2n+1})]^{3.000}~.
\end{equation}
Note that this power law is calculated using data points with $\ln[\mathfrak{R}(\Phi_{2n+1})] \le -7$.  Also note that there appears to be another power law relationship between $\mathfrak{R}(\Phi_{2n-1})$ and $\mathfrak{R}(\Phi_{2n+1})$ that holds for $\mathfrak{R}(\Phi_{2n+1}) \gg 1$ that we do not investigate further in this paper.

We can analytically justify the main features of the fitting formula (\ref{eq:R fit}) as follows:  First, we can solve (\ref{eq:R-Phi ODE}) under the assumption that
\begin{equation}\label{eq:R approx}
	\mathfrak{R} \ll \cos^{2}(\tfrac{1}{2}\Phi)~.
\end{equation}
We would expect this condition to hold for some interval between successive recollapses $\Phi \in [\Phi_{2n-1}+\varepsilon,\Phi_{2n+1}-\varepsilon]$ when the curvature is small $\mathfrak{R} \ll 1$.  Here, we can take $\varepsilon = \mathcal{O}(1) \times \sqrt{\mathfrak{R}(\Phi_{2n+1})}$ since the curvature in the interval is bounded from above by $\mathfrak{R}(\Phi_{2n+1})$.  Making use of (\ref{eq:R approx}) in (\ref{eq:R-Phi ODE}), we obtain
\begin{equation}\label{eq:R change}
	\frac{\mathfrak{R}(\Phi_{2n+1}-\varepsilon)}{\mathfrak{R}(\Phi_{2n-1}+\varepsilon)} \simeq \left[ \frac{1+\cos(\tfrac{1}{2} \varepsilon) }{1-\cos(\tfrac{1}{2}\varepsilon)} \right]^{2/3} \approx \left( \frac{\varepsilon}{4} \, \right)^{-4/3}~.
\end{equation}
The last approximation comes from assuming $\varepsilon \ll 1$ which is consistent with $\mathfrak{R}(\Phi_{2n+1}) \ll 1$.

We now use the approximate solution (\ref{eq:exact 1}) to determine the relation between the spatial curvature on either side of the recollapse at $\Phi=\Phi_{2n-1}$; i.e., $\mathfrak{R}(\Phi_{2n-1}-\varepsilon)$ and $\mathfrak{R}(\Phi_{2n-1}+\varepsilon)$.  We first use (\ref{eq:R-eta-xi}) to re-write (\ref{eq:exact 1}) as
\begin{equation}\label{eq:R cubic}
	0 \simeq \mathfrak{R}^{3} - 4k(3\mathfrak{R} +4\xi^{2})^{2}~,
\end{equation}
where
\begin{equation}\label{eq:asymptotic xi}
	\xi \simeq \tfrac{1}{2}\sqrt{\mathfrak{R} + \tfrac{1}{4}(\Phi - \Phi_{2n-1})^{2}} + \tfrac{1}{4} (\Phi - \Phi_{2n-1}) \simeq \begin{cases} 0, & \sqrt{\mathfrak{R}} \ll \Phi_{2n-1} - \Phi, \\ \tfrac{1}{2} (\Phi - \Phi_{2n-1}), & \sqrt{\mathfrak{R}} \ll \Phi - \Phi_{2n-1},   \end{cases}
\end{equation}
and $k>0$ is a constant.  Equation (\ref{eq:R cubic}) is a cubic equation for $\mathfrak{R}$ that can be solved explicitly.  However, series expansions for $\mathfrak{R}$ prove to be more useful; these are
\begin{subequations}
\begin{align}
	\mathfrak{R} & \simeq 36k \left[ 1 + \frac{2\xi^{2}}{27k} + \mathcal{O} \left( \frac{\xi^{4}}{k^{2}} \right) \right], \\
	\mathfrak{R} & \simeq 4k^{1/3}\xi^{4/3} \left[ 1 + \frac{2k^{1/3}}{\xi^{2/3}} + \mathcal{O} \left( \frac{k^{2/3}}{\xi^{4/3}} \right) \right],
\end{align}
\end{subequations}
for small and large $\xi$, respectively.  Combining these with (\ref{eq:asymptotic xi}) we obtain the approximations
\begin{equation}
	\mathfrak{R}(\Phi_{2n-1} - \varepsilon) \simeq 36k~, \quad \mathfrak{R}(\Phi_{2n-1}+\varepsilon) = 2^{2/3}k^{1/3}\varepsilon^{4/3}~,
\end{equation}
which are valid for $\sqrt{k} \ll \varepsilon \ll 1$.  These in turn imply
\begin{equation}
	9 [\mathfrak{R}(\Phi_{2n-1}+\varepsilon)]^{3} \simeq \varepsilon^{4} \mathfrak{R}(\Phi_{2n-1} - \varepsilon) ~.
\end{equation}
Using this to eliminate $\varepsilon$ from the righthand side of (\ref{eq:R change}), we obtain
\begin{equation}
	\mathfrak{R}(\Phi_{2n-1}-\varepsilon) \simeq \tfrac{9}{256} [\mathfrak{R}(\Phi_{2n+1}-\varepsilon)]^{3}~.
\end{equation}
The $\varepsilon \rightarrow 0$ limit of this is
\begin{equation}\label{eq:R analytic}
	\mathfrak{R}(\Phi_{2n-1}) \sim 0.03516 \times [\mathfrak{R}(\Phi_{2n+1})]^{3}~.
\end{equation}
This is broadly consistent with the result obtained from simulations (\ref{eq:R fit}) in that both formulae predict the same power law $\mathfrak{R}(\Phi_{2n-1}) \propto[\mathfrak{R}(\Phi_{2n+1})]^{3}$, but with different constants of proportionality.

The relations \eqref{eq:R fit} or (\ref{eq:R analytic}) give the approximate evolution of the spatial curvature through the sequence of bounces and recollapses in the perturbative regime $\mathfrak{R}\ll1$.  Both are clearly in agreement with the fact that the $\mathfrak{R}=0$ separatrix is a past attractor for the system.  The situation in the lower half-plane $U<0$ is simply the reverse of what occurs in the $U>0$ half-plane; for $U<0$, the separatrix is approached in the asymptotic future in exactly the same manner.

\subsection{Duration of a single early time bounce-recollapse cycle}
\label{Sec:CycleDuration}

The duration of a bounce-recollapse cycle is related to the spatial curvature.  Defining $\Delta T_{2n+1}$ as the time that elapses between the bounce at $\Phi = \Phi_{2n}$ and the bounce at $\Phi_{2n+2}$, from the first relation in Eq.~\eqref{eq:ds 2D} and from Eq.~\eqref{eq:R def},
\be\label{eq:DeltaT}
\Delta T_{2n+1}=\int_{\Phi_{2n}}^{\Phi_{2n+2}} \de\Phi\;  \frac{1}{6 \cos^2 (\tfrac{1}{2} \Phi ) + 4 \, \mathfrak{R}(\Phi)}~.
\ee
In the early time limit, we have $\mathfrak{R}(\Phi) \ll 1$.  This means that the integral will be dominated by the contribution near the recollapse point $\Phi \sim \Phi_{2n+1}$.  We can therefore approximate it by replacing the limits with $\pm\infty$ and expanding the denominator of the integrand in a series about $\Phi = \Phi_{2n+1}$.  We retain terms up to second order:
\begin{subequations}\label{eq:second order}
\begin{align}
	\cos^2 (\tfrac{1}{2} \Phi ) & \approx \tfrac{1}{4} (\Phi-\Phi_{2n+1})^{2}, \\ \mathfrak{R}(\Phi) & \approx \mathfrak{R}(\Phi_{2n+1}) + \tfrac{1}{2} \sqrt{\mathfrak{R}(\Phi_{2n+1})} (\Phi-\Phi_{2n+1}) + \tfrac{3}{16} (\Phi-\Phi_{2n+1})^{2}~.\label{eq:second order 2}
\end{align}
\end{subequations}
Equation (\ref{eq:second order 2}) is obtained by inserting the differential equation (\ref{eq:R-Phi ODE}) into the series expansion of $\mathfrak{R}(\Phi)$.  We then have
\begin{align}\nonumber
\Delta T_{2n+1} & \approx 4 \int_{-\infty}^{\infty} \frac{\de\Phi}{ 16 \mathfrak{R}(\Phi_{2n+1}) + 8 \sqrt{\mathfrak{R}(\Phi_{2n+1})} (\Phi-\Phi_{2n+1}) + 9 (\Phi-\Phi_{2n+1})^{2} }\\ &  = \frac{\pi}{4} \sqrt{ \frac{2}{\mathfrak{R}(\Phi_{2n+1})} } \nonumber \\ & \approx  1.111 \times [\mathfrak{R}(\Phi_{2n+1})]^{-1/2}. \label{eq:Delta T analytic}
\end{align}

\begin{figure}
\begin{center}
\includegraphics[width=0.8\textwidth]{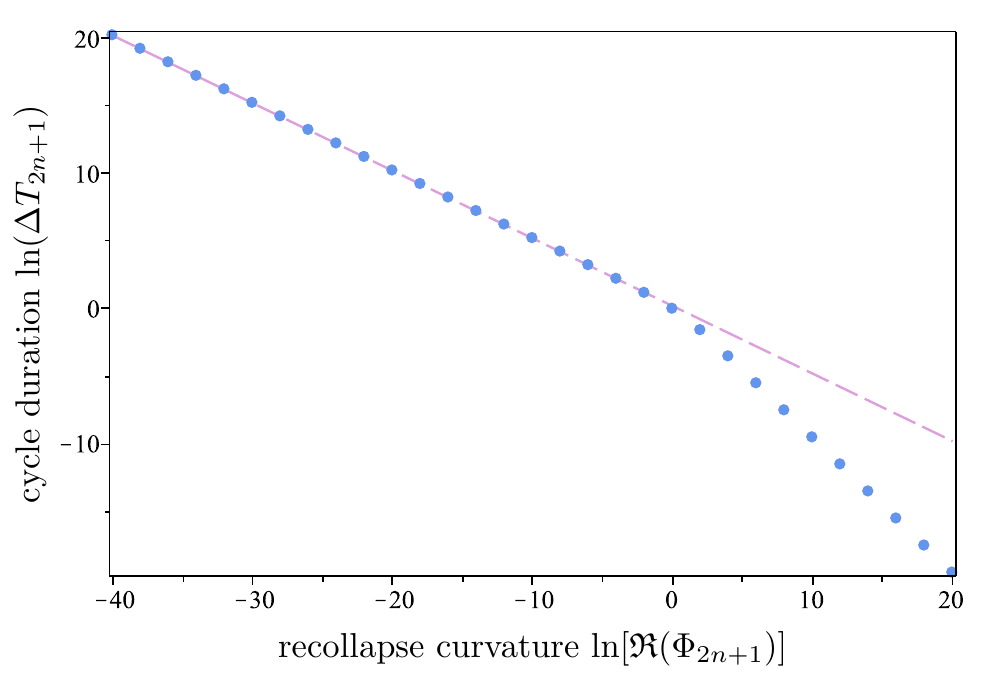}
\caption{Numerical results for the time $\Delta T_{2n+1}$ for one cycle (bounce $\Phi_{2n}$ to bounce $\Phi_{2n+2}$).  The dashed line shows the best fit (\ref{eq:Delta T numeric}) to the simulation data for cycles where the curvature is small; i.e.\ $\mathfrak{R}(\Phi_{2n+1}) \ll 1$.  We do not show the anaytic approximation (\ref{eq:Delta T analytic}) because it is visually indistinguishable from the best fit line on this plot.}\label{fig:DeltaT}
\end{center}
\end{figure}

The duration of each cycle can also be calculated by using the numeric solution of (\ref{eq:R-Phi ODE}) in the integral (\ref{eq:DeltaT}).  We show the result of this computation in Figure~\ref{fig:DeltaT}.  For small values of the curvature at the recollapse point $\mathfrak{R}(\Phi_{2n+1}) \ll 1$, we find the following empirical relationship
\begin{equation}\label{eq:Delta T numeric}
	\Delta T_{2n+1} = 1.208 \times [\mathfrak{R}(\Phi_{2n+1})]^{-0.5000}~.
\end{equation}
Note that this power law is calculated using data points with $\ln[\mathfrak{R}(\Phi_{2n+1})] \le -6$.  Equation (\ref{eq:Delta T numeric}) is similar to the analytic result (\ref{eq:Delta T analytic}).

Both equations (\ref{eq:Delta T analytic}) and (\ref{eq:Delta T numeric}) show that as the spatial curvature becomes smaller and smaller in the past, the duration of the bounce-recollapse cycles gets longer and longer.

\subsection{Early time limit of the radial scale factor $a$}\label{sec:a limit}

In this appendix, we demonstrate that the directional scale factor $a(T)$ approaches 0 as $T \rightarrow -\infty$.  As discussed in Section~\ref{s.dyn-ab}, it is immediate that the limit of $a$ exists, since $a$ is bounded below by 0, and $a$ is monotonically decreasing as $T \to -\infty$ since $H_a>0$.  As a result, to find the limit of $a$ it is sufficient to determine the limit of one subsequence $a(T_i)$, with $\lim_{i\to\infty}T_i = -\infty$.

To do this, it is helpful to first study the subsequence of $S$ evaluated at the bounce points $\Phi=\Phi_{2n}$.  Since $S'/S = \tfrac{1}{2}\sin\Phi$, from the relations \eqref{eq:ds 2D} and \eqref{eq:R def} it is possible to evaluate the relative amplitudes of the value of $S$ at consecutive bounces $\Phi=\Phi_{2n}$ and $\Phi=\Phi_{2n-2}$ with the integral
\be \label{Integral_MinimaS}
\ln\left[ \frac{S(\Phi_{2n-2})}{S(\Phi_{2n})}\right] = -\frac{1}{4}\int_{\Phi_{2n-2}}^{\Phi_{2n}}\de \Phi\; \frac{\sin\Phi}{3\cos^2(\frac{\Phi}{2}) + 2 \mathfrak{R}(\Phi)}~.
\ee
Note that absolute values are not necessary in the logarithm since $S$ is by definition positive.  By splitting the integral into two pieces, the first over the range $\Phi \in [\Phi_{2n-2}, \Phi_{2n-1}]$ and the second over the range $\Phi \in [\Phi_{2n-1}, \Phi_{2n}]$, and introducing the integration variable $\phi = \Phi - \Phi_{2n-2}$ in the first and $\phi = \Phi_{2n} - \Phi$ in the second, this integral becomes
\be
\!\!\! \ln\left[\frac{S(\Phi_{2n-2})}{S(\Phi_{2n})}\right]
=-\frac{1}{2} \int_{0}^{\pi} \!\! \de \phi\; \frac{\sin\phi ~ [\mathfrak{R}(\Phi_{2n}-\phi) - \mathfrak{R}(\Phi_{2n-2}+\phi)]} {\left[ 3\cos^2(\frac{\phi}{2}) + 2 \mathfrak{R}(\Phi_{2n-2}+\phi) \right] \left[ 3\cos^2(\frac{\phi}{2}) + 2 \mathfrak{R}(\Phi_{2n}-\phi)\right]}~.
\ee
Since $\mathfrak{R}(\Phi)$ is an increasing function of $\Phi$, the integrand is positive and therefore
\be
\ln\left[ \frac{S(\Phi_{2n-2})}{S(\Phi_{2n})}\right] \le 0 ~.
\ee
Since the sequence $S(\Phi_{2n})$ is decreasing as $n \to -\infty$ and is bounded below by 0, the limit $\lim_{n \to -\infty}S(\Phi_{2n})$ necessarily exists and is finite.  Furthermore, by \eqref{R->0} $\mathfrak{R}$ vanishes in this same limit and therefore, from \eqref{def-ab}, $\lim_{n \to -\infty}a(\Phi_{2n})=0$.  And since the limit $\lim_{T \to -\infty}a$ exists, as shown above, it must equal the limit of this particular subsequence; therefore%
\footnote{As an aside, by following an analogous argument it is possible to show that the value of the mean scale factor $S$ at the recollapse points $\Phi=\Phi_{2n+1}$ diverges in the limit $n \to -\infty$.  Therefore, $\lim_{T \to -\infty}S$ does not exist.  This behaviour is visible in the righthand panel of Figure~\ref{fig:scalefactor}.},
$\lim_{T \to -\infty}a = 0$.  Also, note that since $\mathfrak{R}$ only vanishes asymptotically, the same is true for $a$: for finite $T$, $a(T) \neq 0$.  

This analysis shows that the solution is non-singular in the past for $U>0$. As the past attractor is approached the curvature invariants oscillate indefinitely and admit no limit. Their upper bound is set by the limiting curvature scale and is given by Eqs.~\eqref{Eq:InvariantsAtBounce} for $\mathfrak{R}=0$, while the lower bound is 0: in the limit $T\to-\infty$, the value of the curvature invariants at the recollapse points gets smaller and smaller, corresponding to a space-time with a larger and larger volume at the recollapse point.

\acknowledgments

We are supported by the National Sciences and Engineering Research Council (NSERC) of Canada and the Atlantic Association for Research in the Mathematical Sciences (AARMS). E.~W.-E.\ is also supported in part by a Harrison McCain Foundation Young Scholars Award. In addition, this research was supported in part by Perimeter Institute for Theoretical Physics. Research at Perimeter Institute is supported by the Government of Canada through Innovation, Science and Economic Development Canada and by the Province of Ontario through the Ministry of Research, Innovation and Science.

\raggedright

\end{document}